# Composite Test inclusive of Benford's Law, Noise reduction and 0-1 Test for effective detection of Chaos in Rotor-Stator Rub

Aman K Srivastava[a], Mayank Tiwari[a], Akhilendra Singh[a]
[a]Indian Institute of Technology Patna, Bihta, Patna , 801106, Bihar

**Abstract** Segregating noise from chaos in dynamic systems has been one of the challenging work for the researchers across the globe due to their seemingly similar statistical properties. Even the most used tools such 0-1 test and Lyapunov exponents fail to distinguish chaos when signal is mixed with noise. This paper addresses the issue of segregating the dynamics in a rotor-stator rub system when the vibrations are subjected to different levels of noise. First, the limitation of 0-1 test in segregating chaos from signal mixed with noise has been established. Second, the underexplored Benford's Law and its application to the vibratory dynamical rotor-stator rub system has been introduced for the first time. Using the Benford's Law Compliance Test (BLCT), successful segregation of not only noise from chaos but also very low Signal to Noise Ratio (SNR) signals which are mainly stochastic has been achieved. The Euclidean Distance concept has been used to explore the scale-invariant probability distribution of systems that comply with Benford's Law to separate chaos from noise. Moreover, for moderate bands of noise in signals, we have shown that the Schreiber's Nonlinear Noise Reduction technique works effectively in reducing the noise without damaging the dynamic properties of the system. Combining these individual layers (0-1 Test, BLCT and Noise reduction) on a rotor system, a Decision Tree based method to effectively segregate noise from chaos and identify the correct dynamics of any system with time series data set has been proposed.

**Keywords** Chaos. 0-1 Test. Benford's Law. Nonlinear Noise Reduction. Composite Test. Rotor Rub.

## 1. Introduction

Over the years, the researchers have tried to understand the nonlinear phenomena in rotating machinery which are responsible for its failure. Rotor-Stator rub is one such phenomena which introduces nonlinear stiffness in the system. This nonlinear stiffness introduces undesired frequencies and vibration in the rotor during operation. The presence of chaotic vibrations in rotor-stator rub was studied in [1] through experimental, numerical and analytical methods. In [2], the presence of chaos in asymmetric rotor with stator rub was studied. Since then, the researchers have been trying to understand the effect of chaos on rotating machinery. Bearing clearance was also identified as a source of chaotic vibration and intermittency was observed in rotor response [3, 4]. The rotor systems supported on oil films also exhibited chaotic vibrations [5, 6]. With the presence of chaos in almost all kinds of nonlinear rotating systems, it became significantly important for researches to identify tools to detect chaos. Lyapunov exponents was one of the first method to be used to detect chaos [7]. Later, it was observed that the presence of noise caused significant deviation in the calculation of Lyapunov exponents [8]. This led to the development of Scale-Dependent Lyapunov Exponents (SDLE) to distinguish noise-mixed chaos from chaos for short time series [9]. The calculation of Lyapunov Exponent was not very cost effective computationally and this led to development of statistical approaches to identify chaos. 0-1 test was introduced, developed, tested and validated in a series of papers [10–14]. Since its development, 0-1 test has found applications in many areas such as finance [15–17], epidemiology [18] and other biological fields [19]. The researchers have used it to detect chaos in Shape Memory Alloy based dynamical systems [20], Strange Non-Chaotic Attractors (SNAs) [21] among others systems [22, 23]. The finite sample study of discrete data sets from financial systems [24] pointed that the 0-1 test failed to detect chaos in noisy data sets. Another study of different types of chaotic systems and noise concluded the misdiagnosis of noise as chaos through 0-1 test [25]. Therefore, it is evident that an effective approach is required to identify the correct dynamics of the system even in the presence of noise. A simple nonlinear noise reduction method proposed in [26] came in handy for removing moderate noise from a limited length data set. The method was widely accepted, modified and applied to time series from different systems [27–29]. Despite this, the identification of chaos in systems with high noise still remains a challenge. Moreover, it is difficult to segregate pure noise from chaos due to their seemingly similarity and effort to reduce noise distorts the dynamics data and makes chaos undetectable.

Most of the naturally occurring sequences, lengths of the river, financial data sets among others are known to follow Benford's Law [30]. This inspired many researchers to explore the idea of compliance of stochastic processes towards Benford's Law [31, 32] and observed scale-invariant property of the probability distribution of systems that comply with Benford's Law. Thereafter, it has been applied to many known fields such as music analysis [33], Geophysical Systems [34, 35]. Nuclear Magnetic Resonance (NMR) signals were observed to follow this law [36]. Researchers investigated the compliance of chaos towards Benford's Law [37] and observed that some of the lower order chaotic systems did while others did not. Moreover, in [32] a few of the chaotic Maps were observed to not comply with the Benford's Law. While the Benford's Law has been applied to many systems, its application to dynamical systems is still not much explored till date. Additionally, its feasibility towards vibration data is still unknown.

The literature survey above, hints towards the lack of effective methods to segregate noise from chaos. A tool which works for all bands of noise in dynamical systems is still unexplored till date. Moreover, the application of known methods such as Benford's Law towards vibratory systems remains to be investigated. Therefore, in this paper we attempt to propose a method that can segregate noise from chaos effectively. On top of that, we have the explored the possibility of identifying the correct dynamics, namely regular (periodic or quasi-periodic) and chaotic dynamics in a rotor-stator dynamical system for all bands of Signal to Noise Ratio (SNR). We have added two layers to the well-known 0-1 test for chaos to segregate and identify all kinds of dynamics of rotor-stator rub model. First layer is Benford's Law Compliance Test (BLCT) and the second layer is Schreiber's Nonlinear Noise Reduction Method. In this study, we numerically integrate the classic rotor-stator rub model [2] to observe periodic, quasi-periodic and chaotic vibrations. Thereafter, we mix the signals with different levels of SNR to check the effectiveness of the proposed method in segregating the noise from chaos and identifying the correct dynamics of the system.

### 2.1 Rotor Rub Model

Figure 1 represents a simple Jeffcott rotor along with stator. The rotor consists of a disk resting on a shaft which is supported by bearings at the two ends. The gap between the rotor and the stator is $\delta$. The contact stiffness between the rotor and the stator is denoted by $K_c$ and the coefficient of friction between the rotor and stator is $\mu$. The mathematical model for rub between rotor and stator is:

$$\begin{aligned} & m\ddot{x} + Kx + C\dot{x} + \frac{K_c(r-\delta)}{r}(x-\mu y)H(r-\delta) \\ & = m\varepsilon n^2\cos(nt) \\ & m\ddot{y} + Ky + C\dot{y} + \frac{K_c(r-\delta)}{r}(y+\mu x)H(r-\delta) \\ & = m\varepsilon n^2\sin(nt) - mg \end{aligned} \qquad (1)$$

Where m is the mass of the rotor, K is the stiffness and C is the damping coefficient. Here, $r=\sqrt{x^2+y^2}$ is the total displacement of the rotor. When r becomes equal to $\delta$, the rubbing starts. H(.) is the Heaviside function and $\varepsilon$ is the eccentricity in the disk. The nonlinear rub contact forces in x and y are represented as

$$F_x = \frac{K_c(r-\delta)}{r}(x-\mu y)H(r-\delta)$$
$$F_y = \frac{K_c(r-\delta)}{r}(y+\mu x)H(r-\delta) \quad (2)$$

**2.2 The 0-1 Test for Chaos**

The 0-1 test [10] segregates regular dynamics from chaotic dynamics in deterministic dynamical systems. The test takes time series data as an input and gives a value of 0 or 1. '0' represents regular dynamics, which is periodic or quasi-periodic vibration whilst '1' represents chaotic vibration. Therefore, the time for post processing of vibration data to find out Lyapunov exponents is saved. The test converts a 1-D time series data into a 2-D system

$$p_c(n+1) = p_c(n) + \phi(n)\cos cn$$
$$q_c(n+1) = q_c(n) + \phi(n)\sin cn \quad (3)$$

Where $c \in (0,2\pi)$ is a constant. Then we define the mean square displacement as

$$M_c(n) = \lim_{N\to\infty}\frac{1}{N}\sum_{j=1}^{N}([p_c(j+N)-p_c(j)]^2+[q_c(j+N)-q_c(j)]^2)$$

n=1,2,3,….. (4)

And the growth rate,

$$K_c = \lim_{N\to\infty}\frac{\log M_c(n)}{\log n} \quad (5)$$

For regular dynamics, $K_c = 0$ and for chaotic dynamics, $K_c = 1$. Due to many issues with choosing a single value of c for the computation of $K_c$ such as resonance as described in [12], a set of values of c are chosen and median of the result is taken. Now, there are two approaches to find out the final value of $K_c$ , regression approach and correlation method.

The correlation method outperforms the regression approach for most of the dynamic systems. This has been established in [12]. Therefore, we use the correlation method to find out the final value of $K_c$ . In the correlation method, two vectors are formed, $\Delta = (M_c(1), M_c(2), \ldots. M_c(N))$ and $\xi = (1,2,3,4,\ldots N)$, and the correlation coefficient between them is calculated as $K_c$.

**2.3 Benford's Law or Law of First Digit**

Benford's Law [30] has known applications in areas such as Fraud Detection, Genome Data, Macroeconomics among others. The Data set that follow Benford's Law has the probability distribution of significant digit defined as

$$P_B = \log_{10}(1+1/d) \quad (6)$$

Where d=1, 2, 3, …9 is the significant digit. Most of the data sets that follow Benford's Law are naturally occurring such as set of numbers in a newspaper or length of rivers in a country. Therefore, it is only natural that stochastic Data sets follow Benford's Law. Presently, no study has been done on vibration data sets from dynamical systems such as Rotor-Stator Rub Problem. The compliance of vibration data sets of different type solutions such as Periodic, Quasi-periodic and Chaotic towards Benford's Law remains a white spot. To measure the compliance, Euclidian distance concept has been used as described in [32]. With the available time series $x_i$, a scaled time series is generated:

$$y_j^s = 1/s \sum_{i=(j-1)s+1}^{js} x_i \;\; , 1 \le j \le [N/s] \quad (7)$$

Where, s=1,2,3,….10 is the scaling factor and **[N/s]** is the greatest integer function. The probability of occurrence of each significant digit in the scaled series ($P_s(d)$) is then evaluated. The Euclidian distance is defined as

$$ED(s) = \sqrt{\sum_{i=1}^{9}(P_{B_i}(d) - P_s(d))^2} \quad (8)$$

The property of noise which will be key in segregating stochastic systems from deterministic systems is that the ED value for noise is almost scale invariant. Moreover, the maximum values of ED at any scale should not cross $\approx 0.1$ideally for pure noise as it complies with the Benford's Law. This test from now onwards will be referred as Benford's Law Compliance Tests (BLCT) in the sections ahead.

## 3. Results

### 3.1 Limitation of 0-1 Test

Many researchers have pointed out clearly that the 0-1 test doesn't computationally distinguish between noise and chaos. In this section, the effectiveness of 0-1 test for Rotor-Stator Rub Mode will be checked. The different types of vibrations such as periodic, quasi-periodic and chaotic Vibration Data will be fed to the 0-1 test and the output will be observed for '0' or '1' value to differentiate regular dynamics from chaotic dynamics. Practically, these different vibration data co-exist with Noise. It is very difficult to acquire a data from any Machine without mixing it with some kind of noise. Therefore, to test the effectiveness of 0-1 test towards real data from machines, we will encode some vibration data with noise.

Parameters given in Table 1, taken from [2], have been used to numerically integrate equation 1 using the varying step size ODE45 Runge-Kutta Method in Matlab©[38]. This has been done to validate the numerical integration method. Figure 2 and 3 in this paper have been generated and are same as Figure 3 and 4 in [2], establishing the correctness of the variable step size ODE45 numerical integration performed here. The static deflection of the rotor has been used as the initial condition for all the simulations. Atleast 500 time periods have been integrated in order to generate a correct Poincare map.

#### 3.1.1 Periodic Case

Set of parameters from Case 1,2 and 3 result in Periodic Vibration of the Rotor-Stator system, as is evident from Figure 2,3 and 4 respectively. Figure 2 clearly points out that the rotor-stator clearance has not been reached and the rotor is exhibiting periodic motion. Figure 3 and 4 represents the initial state of rub between rotor-stator where the orbit is bouncing and hitting the stator but the motion is still periodic. Case 2 results in a period-1 motion while in Case 3 rotor exhibits period-2 motion. The difference is also evident through the frequency spectra which clearly tells that case 3 has subharmonic frequencies in the spectra along with the harmonics. Moreover, Poincare Map of Case 3 has two points implicating period-2 motion.

The important thing to consider while feeding the vibration data from Rotor-Stator system to 0-1 test is that the Poincare points time series data has to be given as input instead of x and y response data. Here, 100 values of c have been chosen between $c \in (0, 2\pi)$ and the median of all the $K_c$ value has been taken to find out the final result of 0-1 test. The 0-1 test on Case 1, 2 and 3 data gives an output close to zero as listed in Table 2 indicating that the motion is non-chaotic.

#### 3.1.2 Quasi-Periodic Case

Set of parameters from Case 4,5 and 6 result in Quasi-periodic motion of the Rotor-Stator System. The conventional method to identify the Quasi-Periodic motion is through the presence of incommensurate frequencies (incommensurate to the rotor operating frequency) in the spectra. Moreover, the points in the Poincare Map align to form a closed loop. For Cases 4, 5 and 6, the frequency spectra are rich with many frequencies between the integers as is evident from Figure 5,6 and 7 respectively. Moreover, in all of the three cases, the Poincare Map has set of points forming a closed loop. Therefore, the rotor is exhibiting Quasi-periodic motion. The orbits in all three cases have toroidal motion.

The Poincare point time series data is fed to the 0-1 test algorithm and the test results are listed in Table 2. The output is close to zero for all the three cases indicating that the motion is non-chaotic.

#### 3.1.3 Chaotic Case

Set of parameters from Case 7,8 and 9 result in chaotic vibration in the rotor-stator system. The frequency spectra of chaotic system have broadband frequency content and a distinctive feature on the Poincare Map which is fractal in nature. Figure 8,9 and 10 shows that the frequency spectra are rich with broadband frequencies apart from the harmonics in the rotor-stator system. The Poincare Map clearly hints towards the chaotic behavior of the system. Furthermore, the 0-1 test has been used to detect the chaotic dynamics in case 7,8 and 9.

The Poincare points generated from case 7,8 and 9, when fed to the 0-1 test algorithm, results in a K value of 1. This indicates the presence of chaotic dynamics for the set of parameters in case 7,8 and 9.

### 3.2 Rotor Response with Noise

The study in previous sections implicates the effectiveness of 0-1 test as a tool for identifying chaos in rotor system. K values from Table 2 vividly segregates the regular dynamics from chaotic dynamics. However, the important thing to investigate is validity of 0-1 test for signal from rotor system mixed noise. To test this, signal from the rotor with regular dynamics (Case 1, 2 and 3) have been mixed with White Gaussian Noise generated in Matlab© [38] using **wgn()** function. White Gaussian Noise has been chosen due to its capability to represent most of the experimentally acquired noise types. SNR has been used to represent the level of noise in signal. The cases for the noisy signal and the results of the 0-1 test for all the cases have been listed in Table 3. Figure 11 (a), (c) and (e) represents the signal from Case 1, 2 and 3 (periodic motion) mixed with white Gaussian Noise with SNR value 50. Figure 11 (b), (d) and (f) shows the corresponding values of $K_c$ which results in a mean value of 1, indicating that the rotor system is exhibiting chaotic dynamics. Therefore, the study of validity of 0-1 test on rotor signal mixed with noise vividly points out the inadequacy of 0-1 test to segregate regular dynamics from chaotic dynamics in the presence of noise in signal.

The first thing that comes to mind to resolve this issue is to use any conventional noise reduction technique. In contrary, the conventional noise reduction methods might affect the dynamics of chaotic system as noise and chaos have similar statistical properties. Moreover, the effectiveness of these techniques also depends on the level on signal mixed with noise. The noise reduction might not be accurate for very low SNR values. To overcome these issues, two things needs to be worked upon. Firstly, we need to have a tool to classify the signal mixed with noise as stochastic or deterministic. There will be SNR values below which the signal will lose any deterministic property and hence that needs to be identified and classified. Secondly, we need to have an effective nonlinear denoising technique which will remove the noise from the system without affecting its dynamics. BLCT as described in the section 2.3 has been used in this study to distinguish stochastic and deterministic processes. Moreover, we have chosen Schreiber's nonlinear noise reduction technique for removing noise from the system.

### 3.3 Benford's Law Compliance Test (BLCT) on Rotor-Stator Rub Data

Over the years, the segregation of noise from chaos has remained a challenge for the researchers due to their similar statistical properties. Section 3.2 established that even 0-1 test could not computationally distinguish between chaos and noise. The segregation of noise from periodic and quasi-periodic vibrations can be performed in many established ways but identifying and separating noise from chaos still remains an obstacle. That is where Benford's Law comes into picture due to its ability to separate any stochastic system from deterministic systems, even chaotic. The input for the BLCT is again the Poincare points. The test isolates the significant digits from the data sets after introducing the scaling as discussed in section 2.3. The Euclidean distance (ED) is then computed for each scale and the variation of ED versus scale determines if the system is stochastic or deterministic. In section 2.3, maximum ED value of $\approx 0.1$was mentioned to be the limit for pure noise but in practical when the signal is mixed with the noise, the actual ED value might be a little higher. But if the signal still has the noise property, it will remain approximately scale invariant. The important factor to consider when deciding the cut off value of ED for signal mixed with noise is to consider the limitation of noise reduction techniques in removing noise when the SNR values are very low. Based on observation, a cut off value of 0.25 is set for ED for cases of signal mixed with noise. The signal with very low SNR remains almost scale invariant until ED values of 0.25 as is observed for both periodic and Quasi-periodic signals. The box containing signals with ED values from 0 to 0.25 and remaining almost scale invariant has been referred to as the bounding box. Figure 12(a) presents the distribution of significant digit in chaotic system (case 7) compared with Benford's Law while Figure 12(b) presents the distribution of significant digit in White Gaussian Noise. It can be clearly observed that the chaotic system deviates from the Benford's Law whilst the noise closely follows the Benford's Law. Figure 13 shows the variation of ED at different scales for White Gaussian Noise and 2 different chaotic signals (case 7 and 9). The result is in accordance with the earlier statement that BLCT can distinguish between noise and chaos. With this established, that BLCT works for chaotic systems as well, it can be applied to different signals mixed with noise to classify them as stochastic or deterministic. Figure 14 shows the ED values at different scales for periodic signal mixed with different levels of noise. As can be seen from the figure, for lower SNR values (till 0.15 SNR) the system is stochastic and after that the ED values are not scale invariant. Figure 15 shows the similar trend for a quasi-periodic system where signal with SNR 1.12 and below behaves like a stochastic system and post that ED starts to vary with scale. This sets up the criteria for identifying stochastic processes from chaotic, periodic and quasi-periodic systems.

### 3.4 Schreiber's Denoising Technique

The method proposed by Schreiber [26] has been used for removing noise from nonlinear systems. The method simply replaces any point in the time series with the average value of this point over points available in a chosen neighborhood. The main advantage of this method over other noise reduction techniques is that it creates a phase plot with the available time series to select the neighborhood and evaluate the average value of the point. The 'k' past coordinates and 'l' future coordinates has been chosen to create embedding vectors for every point in the time series.

$$x_i = (x_{i-k}, \ldots\ldots x_{i+l}) \tag{9}$$

The condition for selecting the neighborhood radius $r_a$ for a point $x_j$ is that the distance of all the trajectory $|x_j - x_i| < r_a$. Utmost care should be taken when selecting the neighborhood radius as a larger value will result in over-denoising and a smaller value might cause under-denoising. Here, the time series input to the Schreiber's technique is the set of Poincare points for a particular operational speed of rotor. In ideal case, the noise reduced time series will have the same points as in the original time series without the noise. But in practical, there will be error depending on the selection of neighborhood radius. To ensure an optimum value of neighborhood radius for least error, signal from case 1 and 6 has been mixed with White Gaussian Noise and denoising has been performed at different values of $r_a$. Figure 16 shows the variation of squared error with $r_a$ for both the cases. The least error is achieved at a $r_a$ value 3 times the rms value of signal amplitude. The signal to noise ratio for both the cases is 30. This optimum value of $r_a$ has been used for all the vibration data sets from rotor model. The effectiveness of any noise reduction method can be measured using two criteria, first being that it should be able to reduce the noise from a noise-mixed signal and second being that it should not disturb the underlying dynamics of the system. Figure 17 shows the effectiveness of Schreiber's technique to reduce noise from a quasi-periodic system (case-6) with a SNR value of 45. The method restores the original dynamics of the system as can be seen in the Poincare Map of the denoised signal in Figure 17(b). Figure 18 shows the Poincare Map of a chaotic signal (case-8) without any noise and the denoised Poincare Map of the same signal. Both the maps are similar. Therefore, it is safe to say that the noise reduction process does not lead to loss of dynamic properties of the system. Now, to show the effectiveness of the noise reduction technique we mix periodic and quasi-periodic signals with different levels of noise. Figure 19 compares the K-values obtained from the 0-1 test for the periodic signal mixed with noise. The result shows that the 0-1 test misdiagnoses the signal mixed with noise but post noise reduction, the test yields a K-value of '0'. Similarly, Figure 20 shows the K-values for quasi-periodic signal mixed with noise. The 0-1 test fails for SNR less than 1000 with the noise in the system but the test after noise reduction process yields a correct K-value of '0'. The important thing to note here is that for very low SNR values, the Schreiber's noise reduction technique leads to the loss of dynamics of the system. Figure 21 presents one such case for quasi-periodic signal. Figure 21(a) shows the Poincare Map of the quasi-periodic signal (case 6) with 0.38 SNR and Figure 21(b) shows the Poincare Map of the signal post noise reduction. Ideally, the Poincare map after noise reduction should have been a closed orbit (Figure7(d)) but it collapses to a single point, which leads to misinterpretation that the system is periodic whilst it originally was quasi-periodic. The reason for such behavior is that the system due to high level of noise has lost its original property and has become stochastic. But fortunately, the BLCT test takes care of it for us, in correctly diagnosing the same signal as stochastic as earlier presented in Figure 15. Hence the tuning between the BLCT and the denoising techniques works like a lock and key mechanism as all the signals for which the denoising technique is not able to restore the original dynamics due to very high level of noise in it, BLCT correctly diagnoses it as stochastic.

## 4. Decision Tree for the proposed Test

Based on the study in previous sections, following steps can be followed to effectively detect chaos in any system:

- Sample the Poincare points from the acquired time series data. Sufficient number of data points should be available in the Poincare data set (at least 500 cycles).
- Perform the 0-1 test for chaos on the Poincare data set and obtain the K-value. If the K-value is 0, the system has regular dynamics. If the K-value turns out to be 1, there are 3 possibilities.
    - The system is purely stochastic
    - The signal is mixed with certain amount of noise
    - The system is actually chaotic.

    So, if the K-value is 1, forward the data to the BLCT algorithm.
- If the ED values are scale invariant and lies inside the bounding box at different scales, the system is stochastic. Otherwise, the system could be either mixed with moderate amount of noise or is actually chaotic. The former issue can be resolved using Schreiber's nonlinear noise reduction technique. This technique will not affect the dynamics of the system as shown in section 3.4.
- Post denoising, perform the 0-1 test for chaos again. This time a K-value of '0' represents regular dynamics whilst a K-value of '1' represents chaotic dynamics.

Figure 22 represents a decision tree flow chart for the process. Table 4 lists the results of the proposed test compared with standard 0-1 test on different cases from rotor-stator rub system mixed with noise. It is evident that the proposed test is able to identify the correct dynamics of the system.

## 5. Conclusion

The limitations of the 0-1 test for chaos and the lack of methods to effectively detect chaos even in noisy systems led to the above study in this paper. Based on the study presented, following can be concluded:

1. The rotor exhibits periodic, quasi-periodic and chaotic motion as presented in section 3. The case 1, 2 and 3 in Table 1 exhibits periodic motion as shown in Figure 2,3 and 4 respectively. The quasi-periodic motion is confirmed by the closed orbits on the Poincare Map in Figure 5(d), 6(d) and 7(d) for case 4, 5 and 6 respectively. The chaotic dynamics in rotor rub model has been observed for case 7, 8 and 9 and confirmed through attractors in Poincare Map (Figure 8(d), 9(d) and 10(d)) and broadband nature of the frequency spectrum (Figure 8(c), 9(c) and 10(c)).

2. The 0-1 test resulted in K-value of '0' when subjected to periodic and quasi-periodic vibrations from rotor-stator rub model (Table 2) while the chaotic cases (case 7, 8 and 9) resulted in a K-value of '1'. But the same cases of periodic and quasi-periodic vibration from rotor when mixed with noise led to misdiagnosis by 0-1 test resulting in a K-value '1' (Table 3). Therefore, it computationally fails to distinguish between chaotic and a regular system with noise.
3. White Gaussian Noise closely follows Benford's Law while chaotic system does not follow Benford's Law as depicted in Figure12. The same can be observed mathematically in Figure13 where the ED value for noise is scale invariant whilst ED values for chaotic systems (case 7 and 9) varies strongly with scale. The Benford's Law Compliance Test confirms that there are certain levels of noise for which a system is stochastic and has lost its original properties. For periodic case, Figure 14 shows that the ED values for case 1 signal with less than 0.15 SNR are scale invariant implying that the system is stochastic. Similarly, Figure 15 shows that for Quasi-periodic system (case 9) mixed with noise, all the signals with SNR value of 1.12 or less are stochastic as the ED values are scale invariant.
4. The Schreiber's nonlinear noise reduction technique is efficiently able to remove noise from any system be it periodic, quasi-periodic or chaotic as depicted in Figure 17. Additionally, it does not lead to any loss in dynamics of chaotic attractors (case 8) as clearly pointed out in Figure 18. The K-values for periodic and quasi-periodic case with varying SNR is shown in Figure 19 and 20, which shows that the proposed test is able to restore the original dynamics of the system. However, there are cases of very high noise, where the denoised Poincare Map is incorrect (Figure 21) implying towards the fact that for some cases of very low SNR, the noise reduction method might not be very efficient. The reason for that being the loss of original signal properties as the system is almost stochastic due to high noise. Fortunately, BLCT takes care of such signals by correctly identifying them as stochastic.
5. The proposed test for chaos in this study adds two layers to the well know 0-1 test. Firstly, the Benford's Law Compliance Test takes care of the signal with very low SNR or pure noise which were originally identified as chaos by the 0-1 test by marking them as stochastic. The second layer, that is the Schreiber's Nonlinear noise reduction method, takes care of the moderate level of noise and restores the original dynamics of the system which had originally regular dynamics but was misdiagnosed as chaotic by the 0-1 test. The misdiagnosed signals in Table 3, after applying the proposed test lists the correct K-values and dynamics of the system in Table 4.
6. The proposed test, although applied here for rotor-stator rub model, is valid for any data set where the noise or the dynamics of the systems are of interest. For example, the test can be applied to weather data sets where chaos is common, heart rates of a patients where chaos leads to heart attacks or in the financial markets where non-compliance to Benford's Law straightaway means a fraud. Hence, the test can be extended to many Biological, economic and financial systems among others.

**Acknowledgement**

The authors would like to acknowledge the financial support provided by MHRD, Govt. of India and General Electric, India.

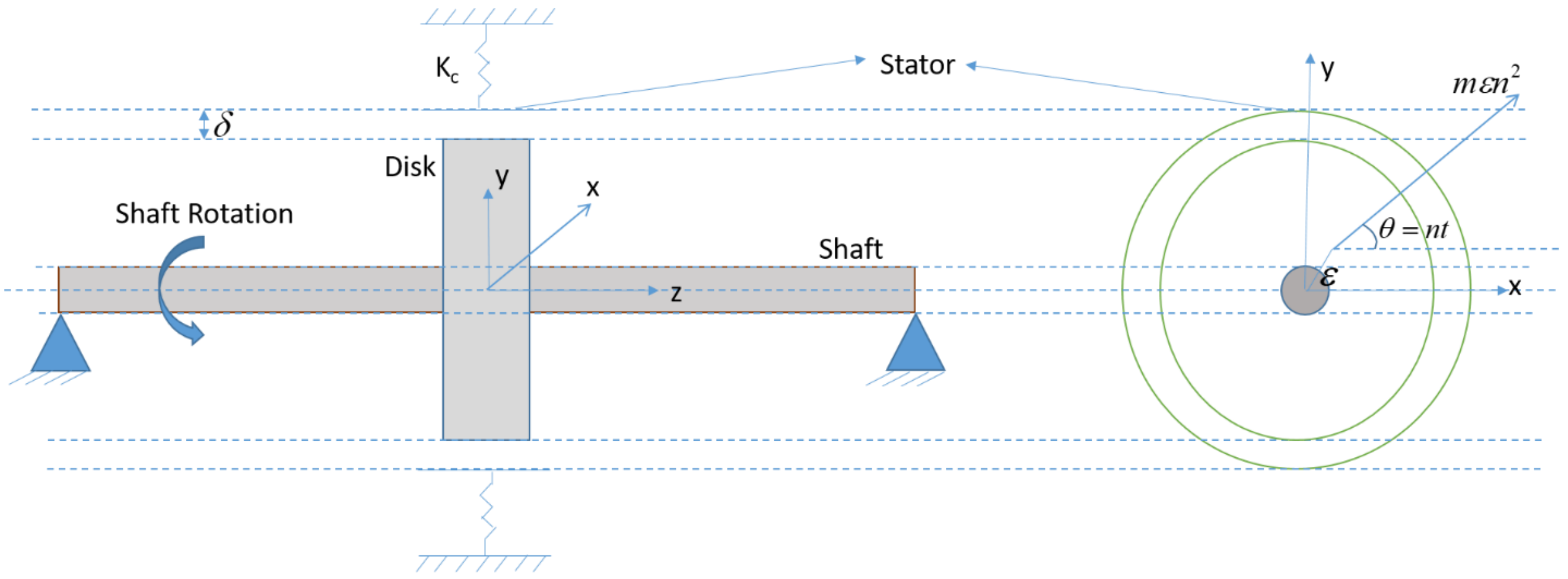

Figure 1. Jeffcott Rotor Model with Stator

| Table 1. Parameters used for Numerical Integration of Rotor-Stator Rub | | | | | | | | |
|---|---|---|---|---|---|---|---|---|
| **Case** | M | K | $\mu$ | $\delta$ | $\varepsilon$ | $K_c$ | C | n |
| **1** | 15 Kg | 2e06 N/m | 0.25 | 9e-05 m | 5e-05 m | 400*K | 3900 Ns/m | $0.5 * \sqrt{Kxx/m}$ |
| **2** | 15 Kg | 2e06 N/m | 0.25 | 9e-05 m | 5e-05 m | 400*K | 3943.6 Ns/m | $1.54 * \sqrt{Kxx/m}$ |
| **3** | 15 Kg | 2e06 N/m | 0.25 | 9e-05 m | 5e-05 m | 400*K | 4820 Ns/m | $1.54 * \sqrt{Kxx/m}$ |
| **4** | 5 Kg | 1e06 N/m | 0.2 | 9e-05 m | 4e-05 m | 400*K | 1341.6 Ns/m | $2.25 * \sqrt{Kxx/m}$ |
| **5** | 5 Kg | 1e06 N/m | 0.2 | 9e-05 m | 4e-05 m | 400*K | 1341.6 Ns/m | $2.41 * \sqrt{Kxx/m}$ |
| **6** | 5 Kg | 1e06 N/m | 0.2 | 9e-05 m | 4e-05 m | 400*K | 1341.6 Ns/m | $2.35 * \sqrt{Kxx/m}$ |
| **7** | 15 Kg | 2e06 N/m | 0.25 | 9e-05 m | 5e-05 m | 400*K | 4309.5 | $1.54 * \sqrt{Kxx/m}$ |
| **8** | 5 Kg | 1e06 N/m | 0.2 | 9e-05 m | 4e-05 m | 400*K | 1341.6 Ns/m | $1.34 * \sqrt{Kxx/m}$ |
| **9** | 15 Kg | 2e06 N/m | 0.25 | 9e-05 m | 5e-05 m | 400*K | 4309.5 | $1.7 * \sqrt{Kxx/m}$ |

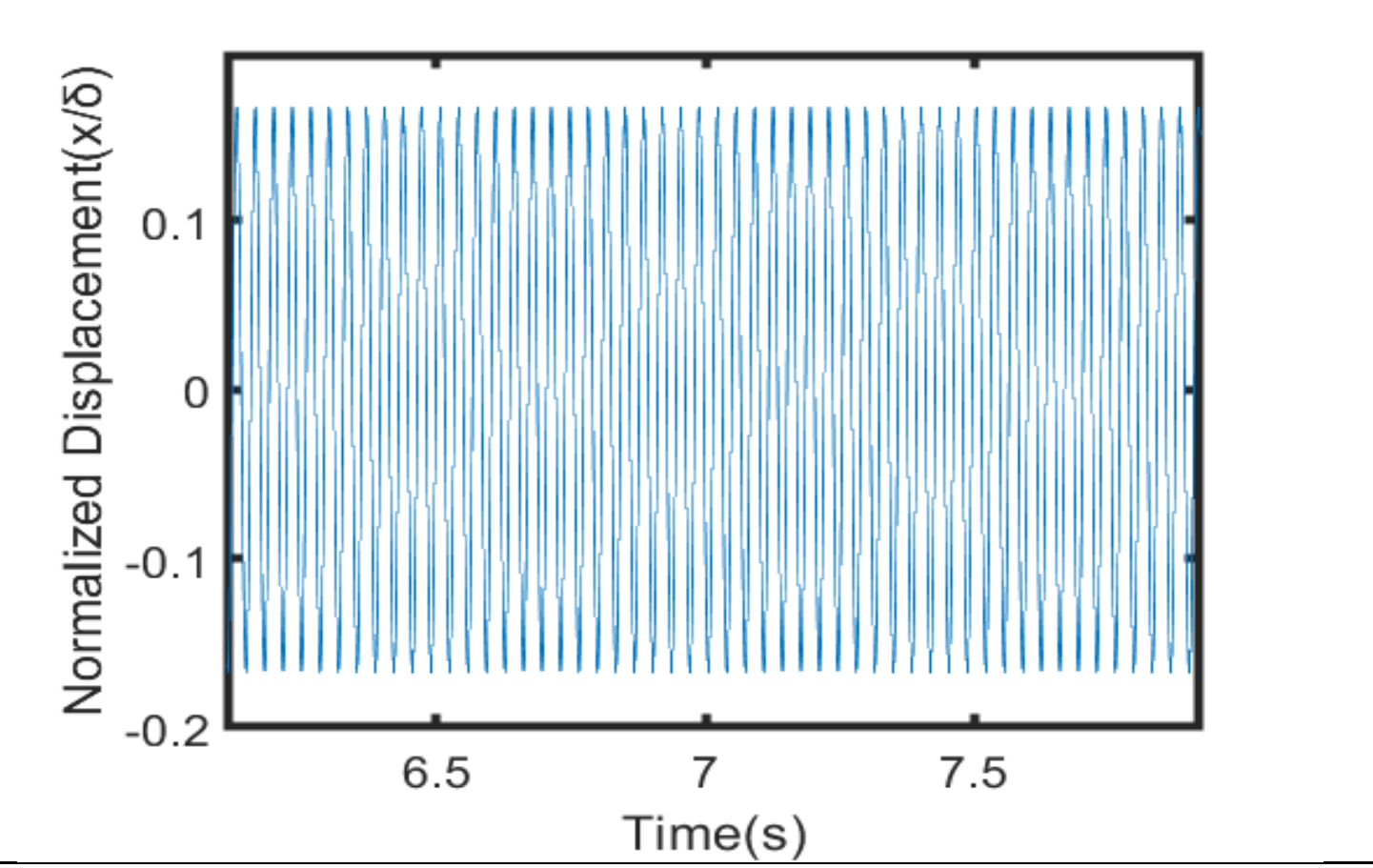

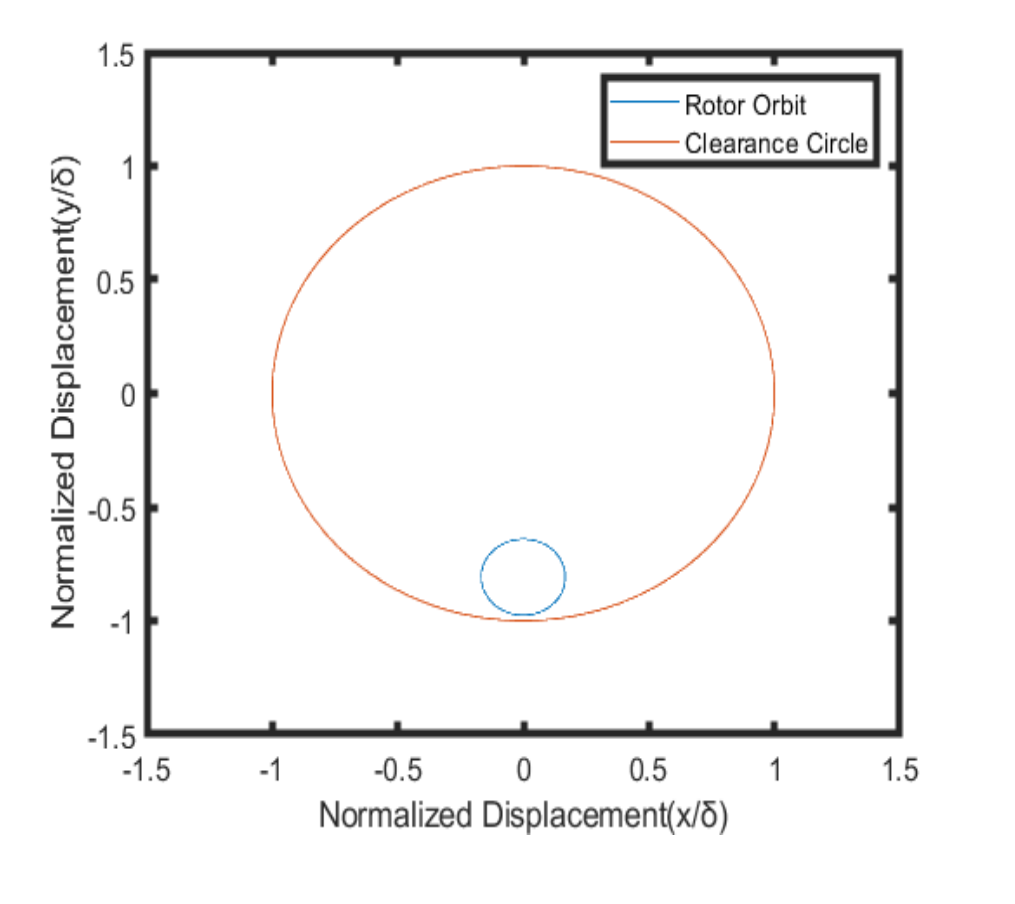

(a) (b)

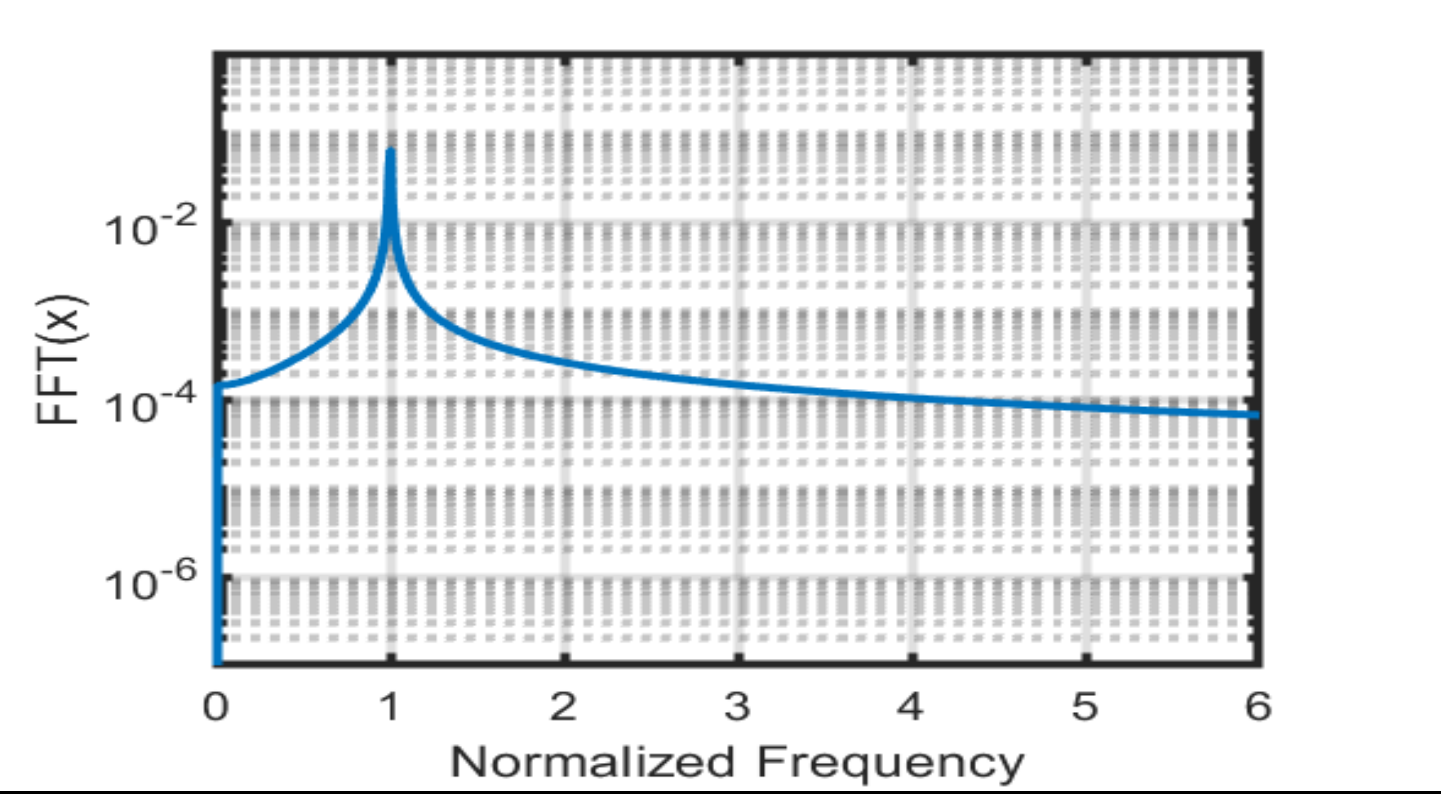

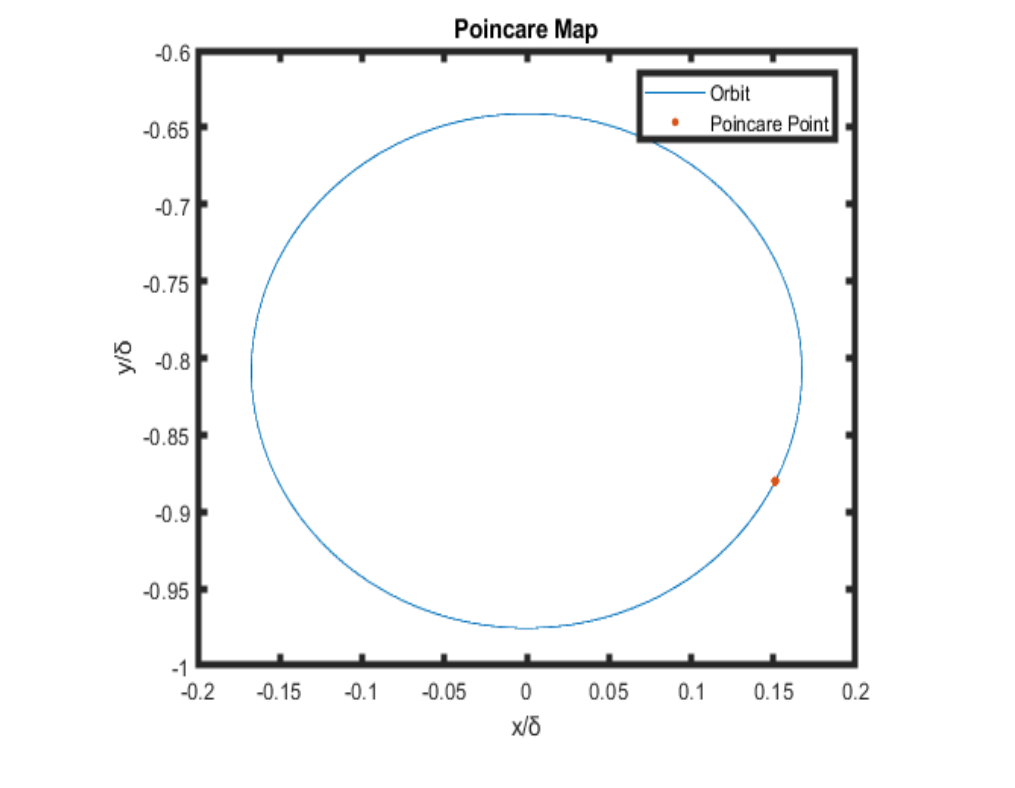

(c) (d)

Fig 2. Rotor Response for case 1 (a) non-dimensional displacement (b) Rotor Orbit (c) FFT (d) Poincare Map

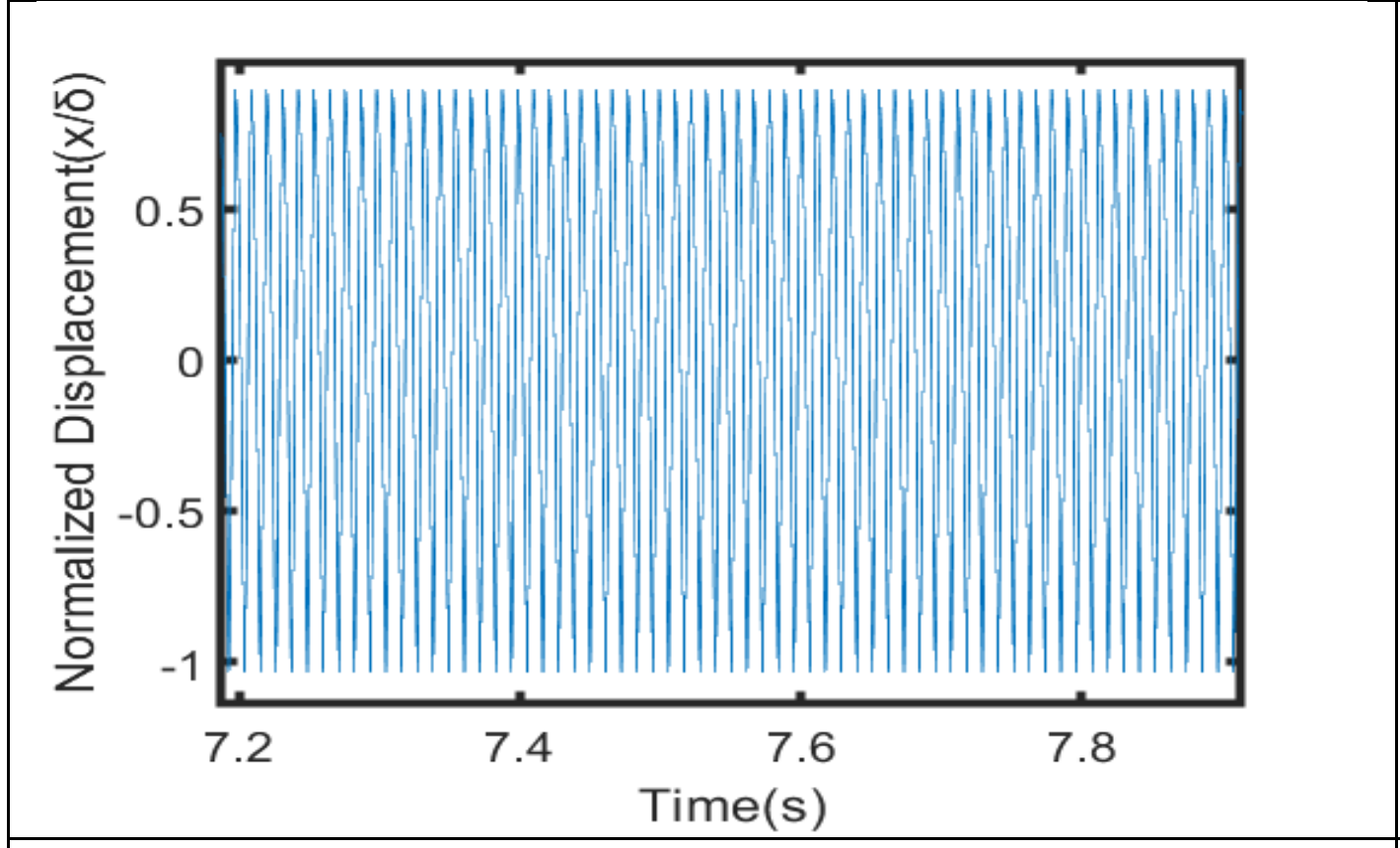

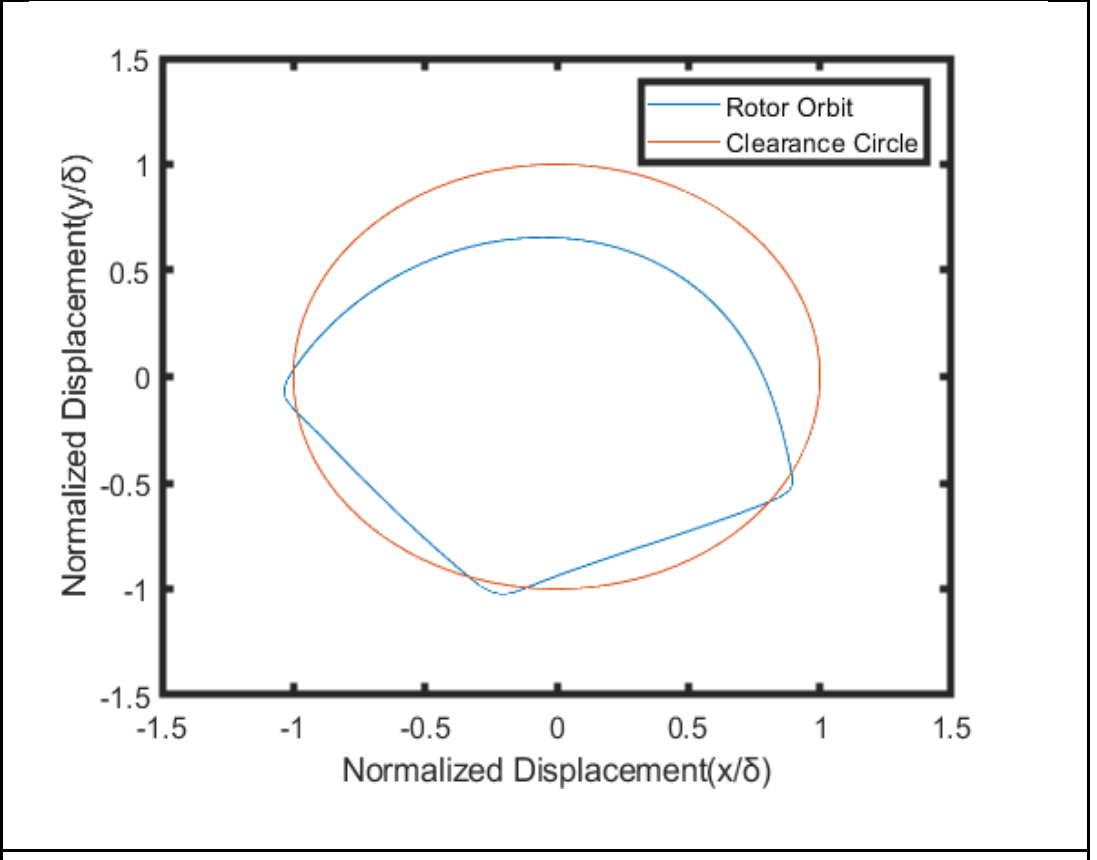

(a) (b)

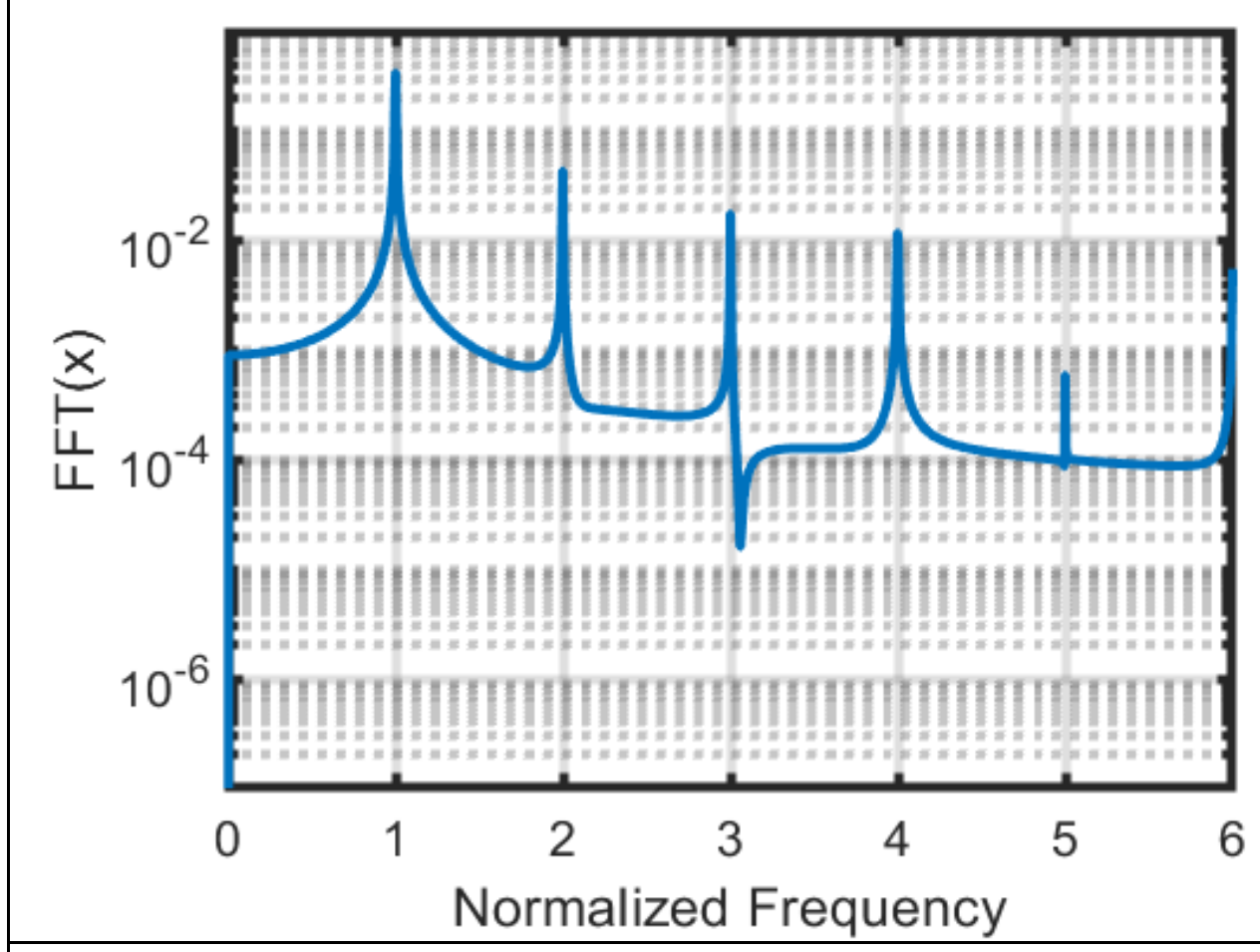

(c)

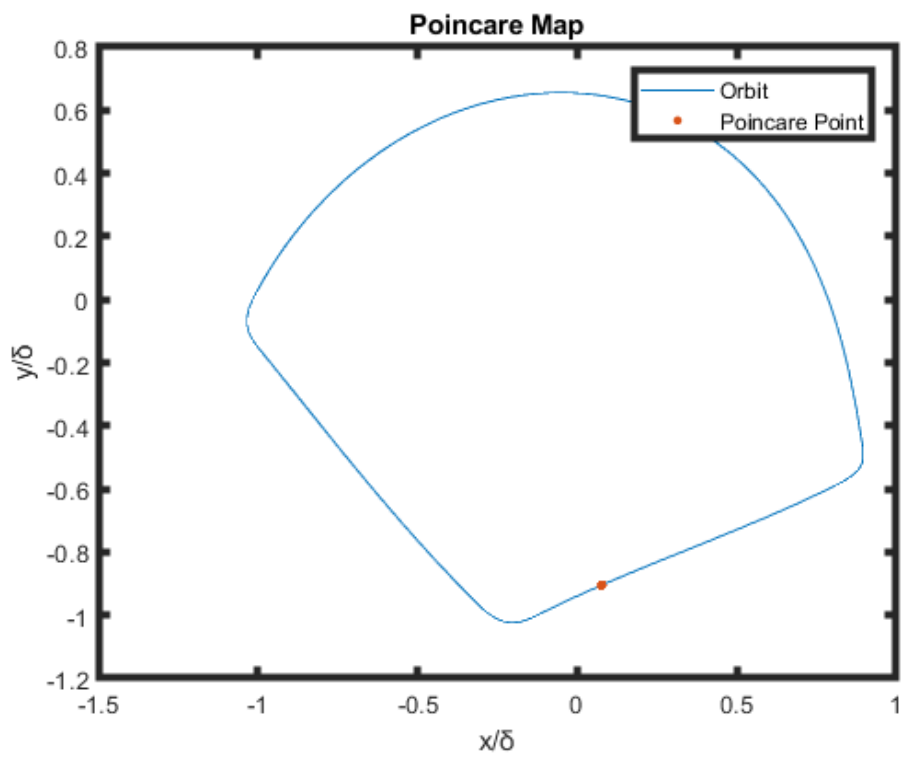

(d)

Fig 3. Rotor Response for case 2 (a) non-dimensional displacement (b) Rotor Orbit (c) FFT (d) Poincare Map

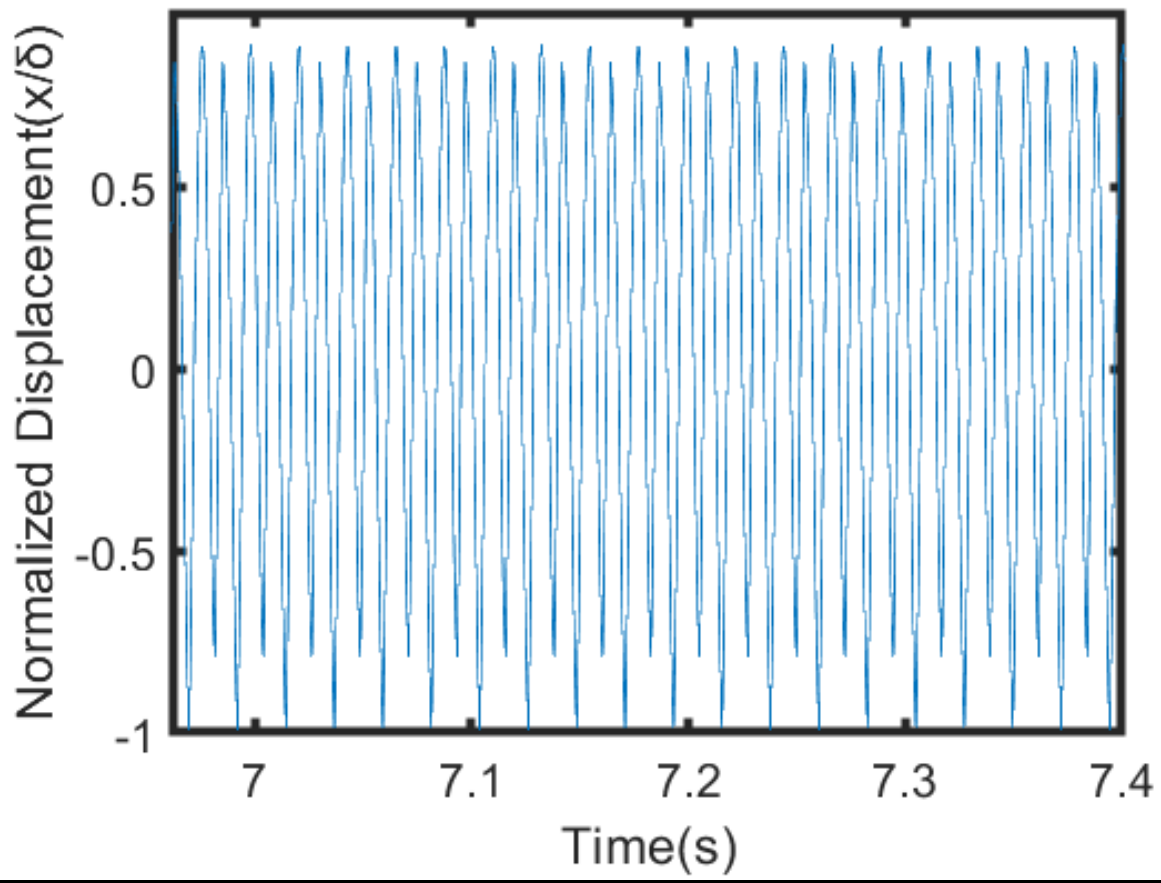

(a)

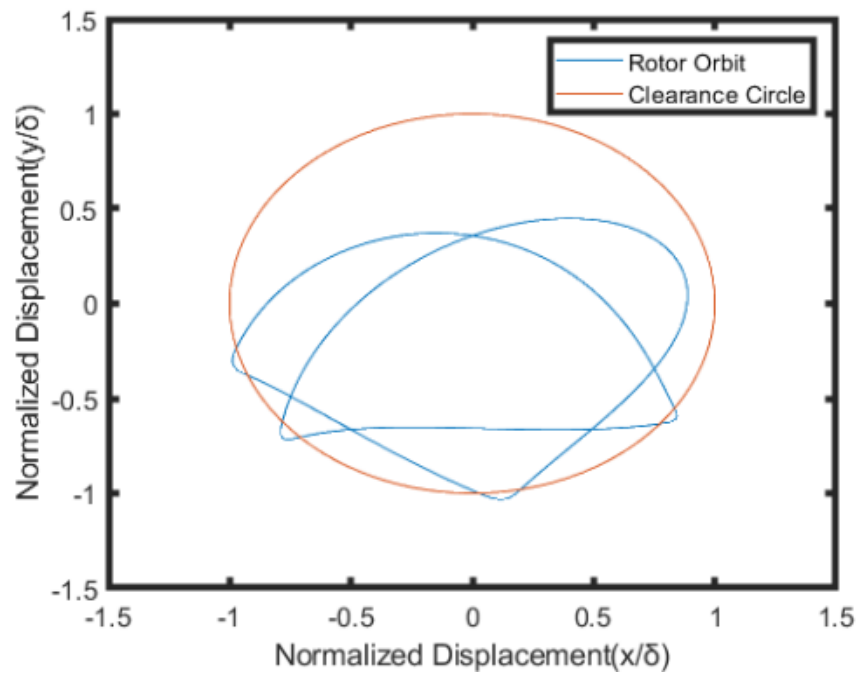

(b)

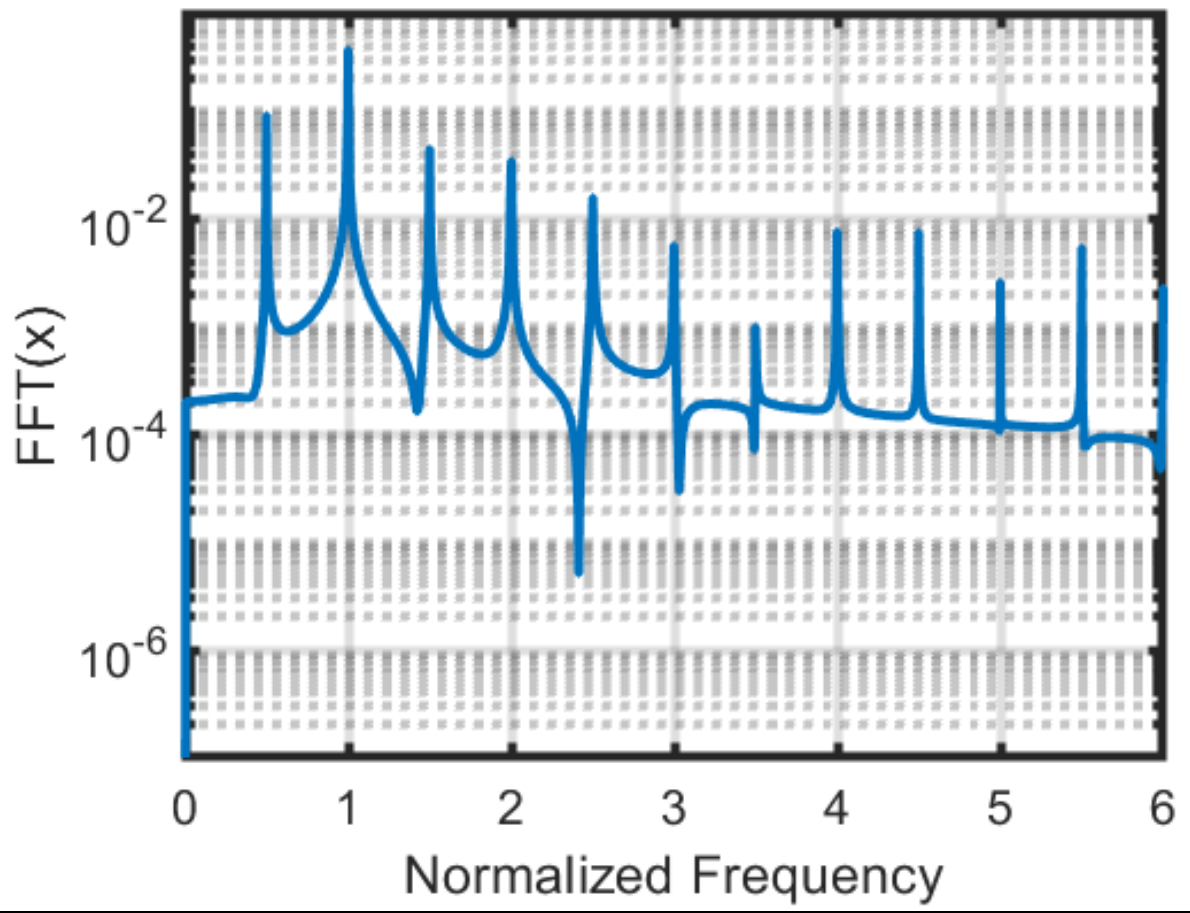

(c)

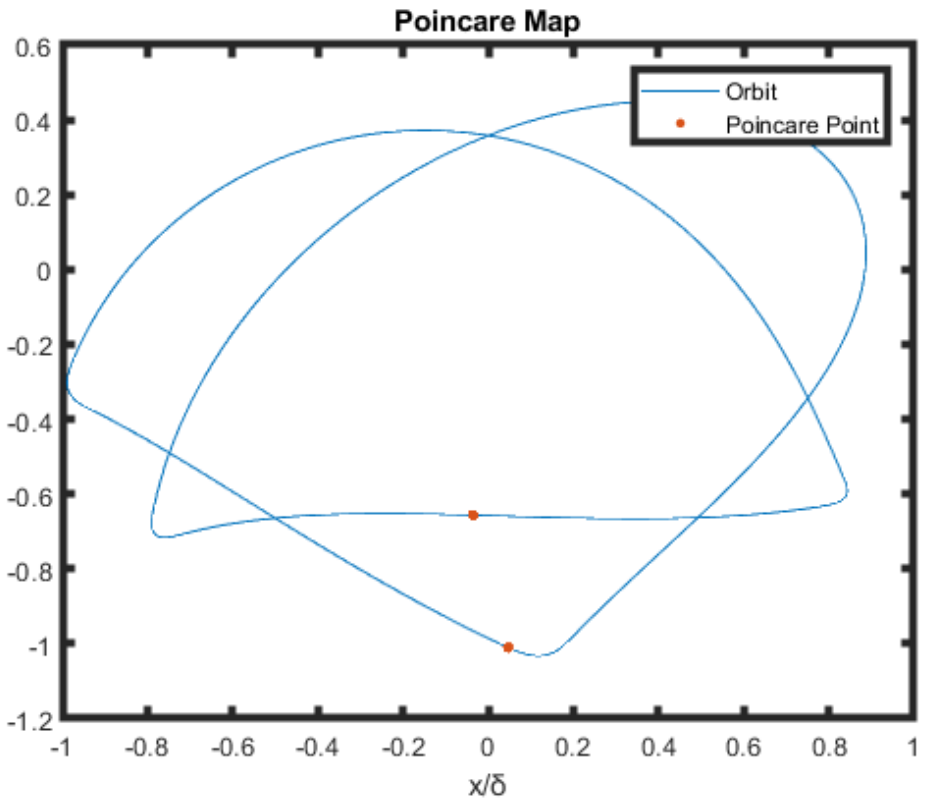

(d)

Fig 4. Rotor Response for case 1 (a) non-dimensional displacement (b) Rotor Orbit (c) FFT (d) Poincare Map

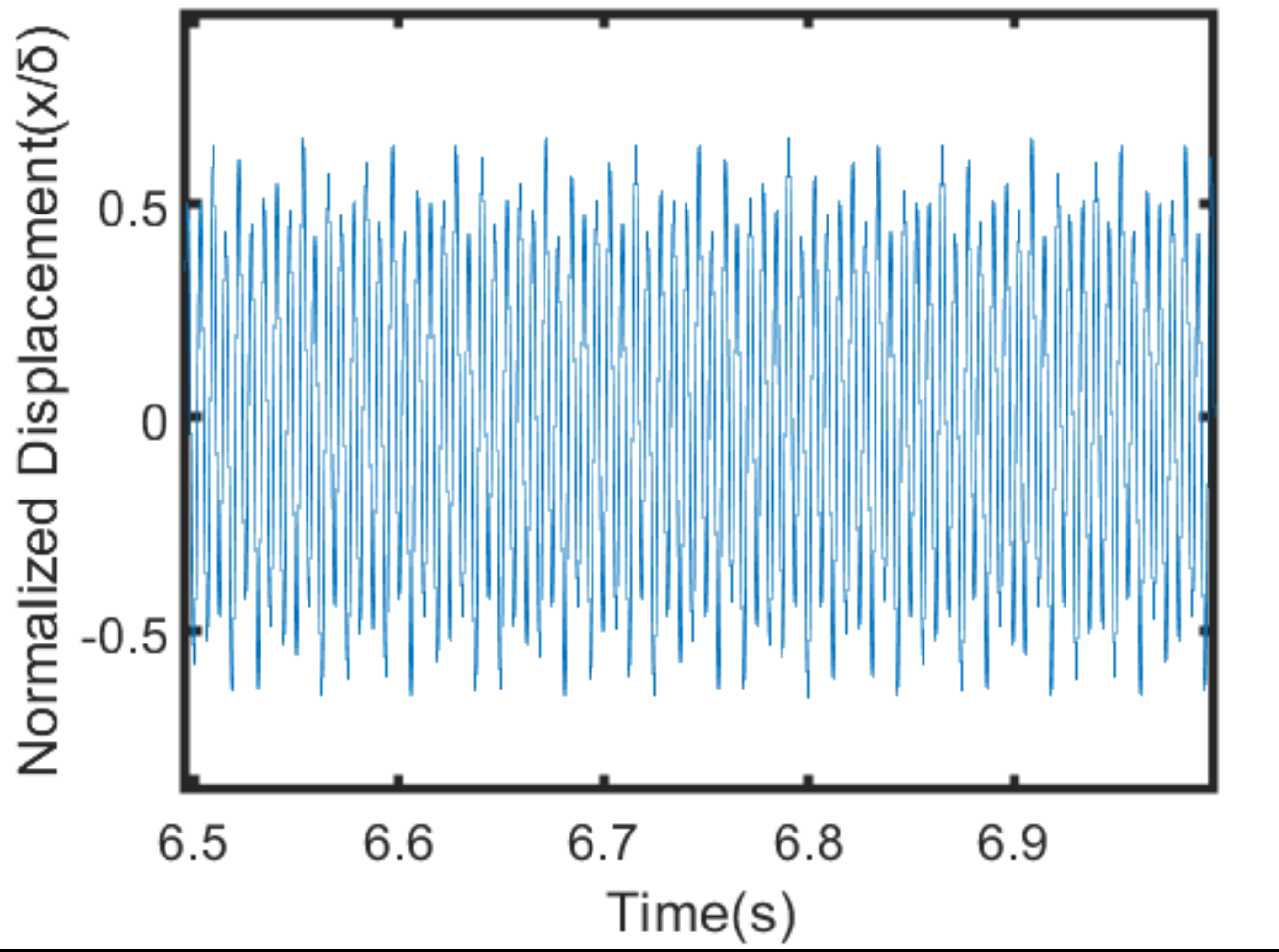

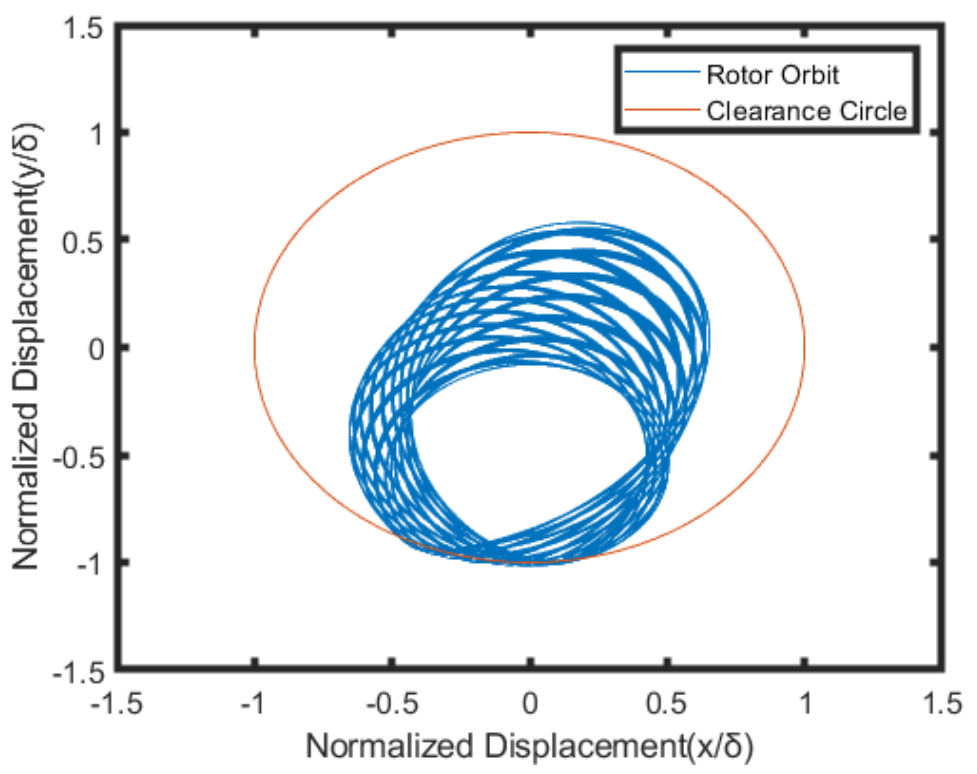

(a) (b)

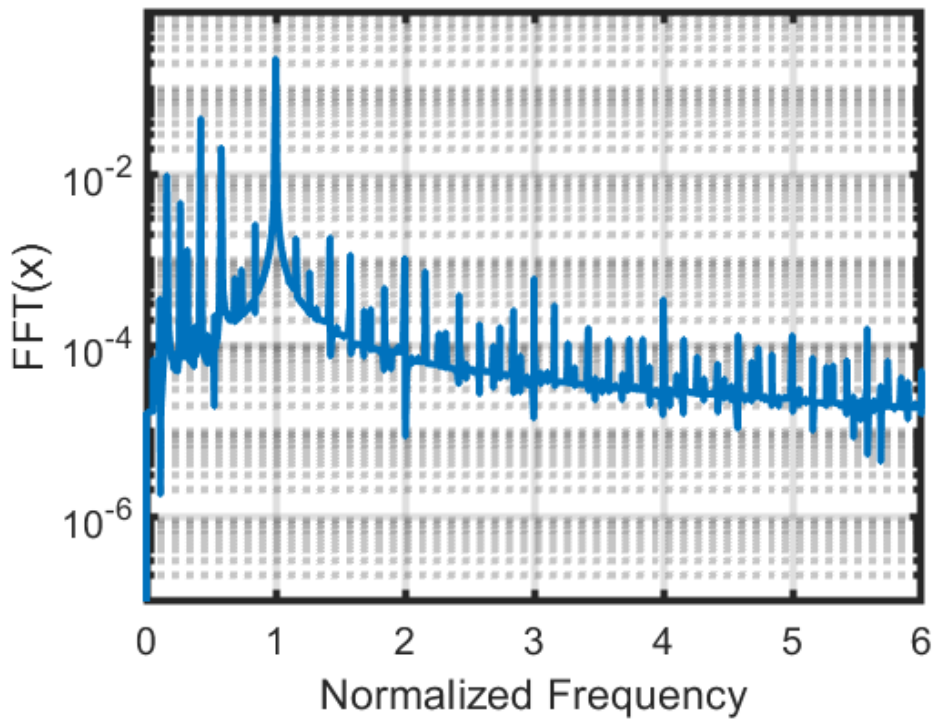

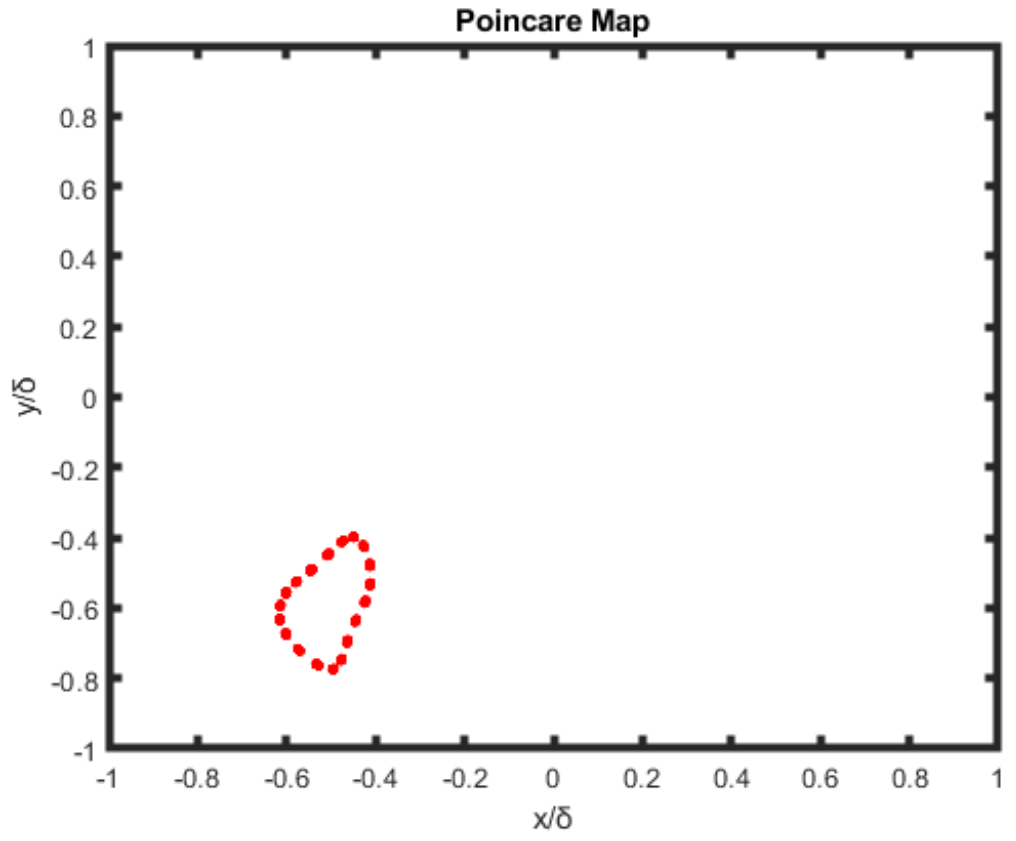

(c) (d)

Fig 5. Rotor Response for case 1 (a) non-dimensional displacement (b) Rotor Orbit (c) FFT (d) Poincare Map

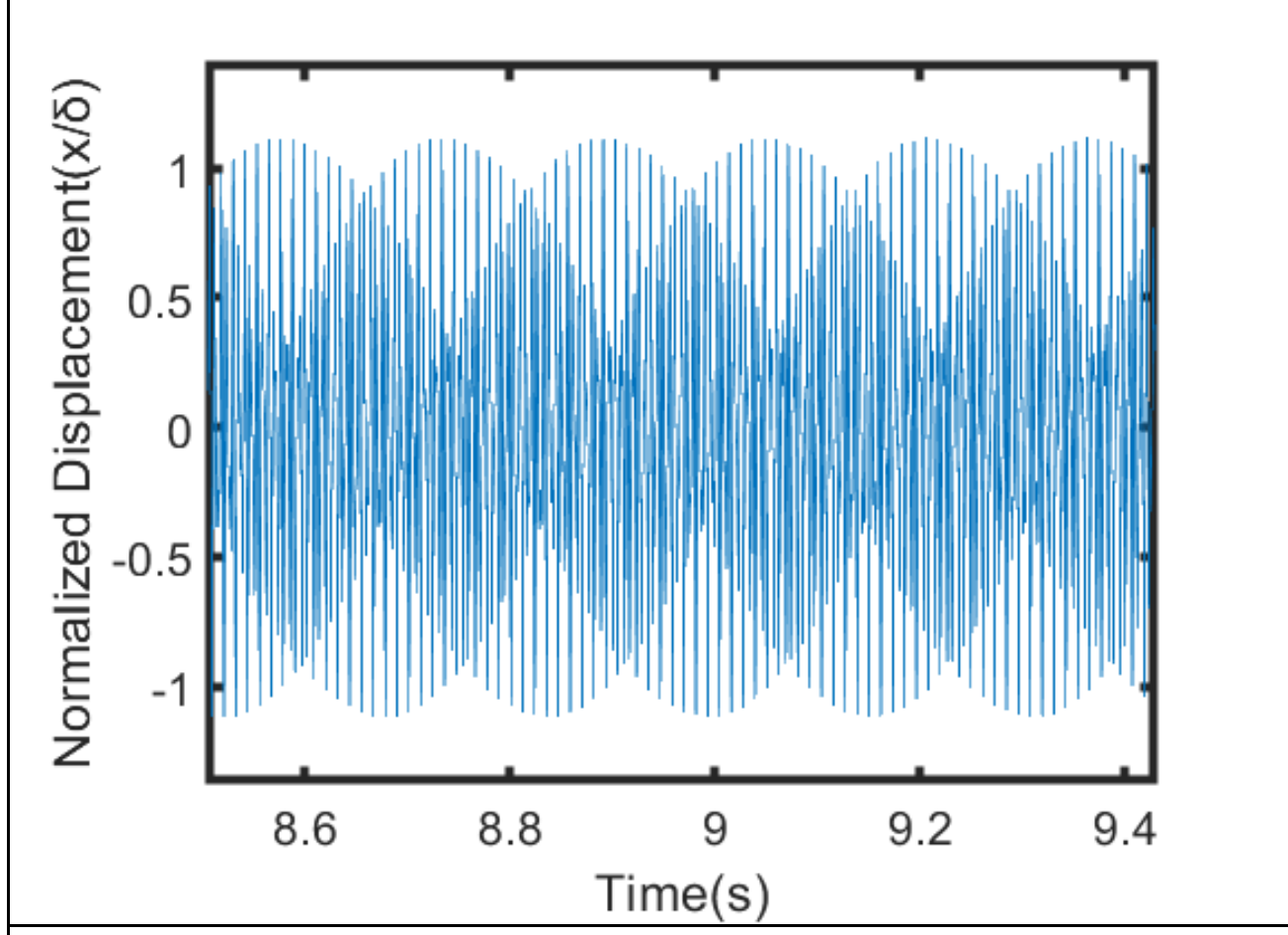

(a)

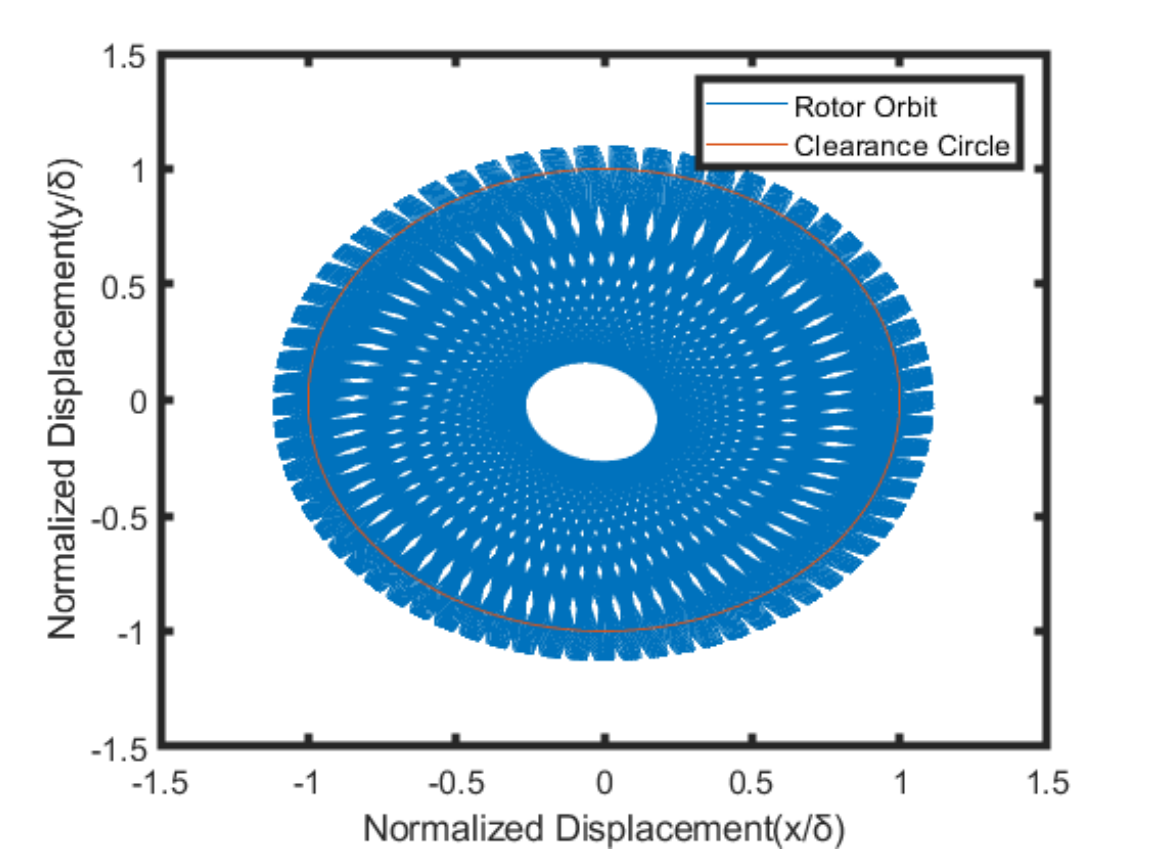

(b)

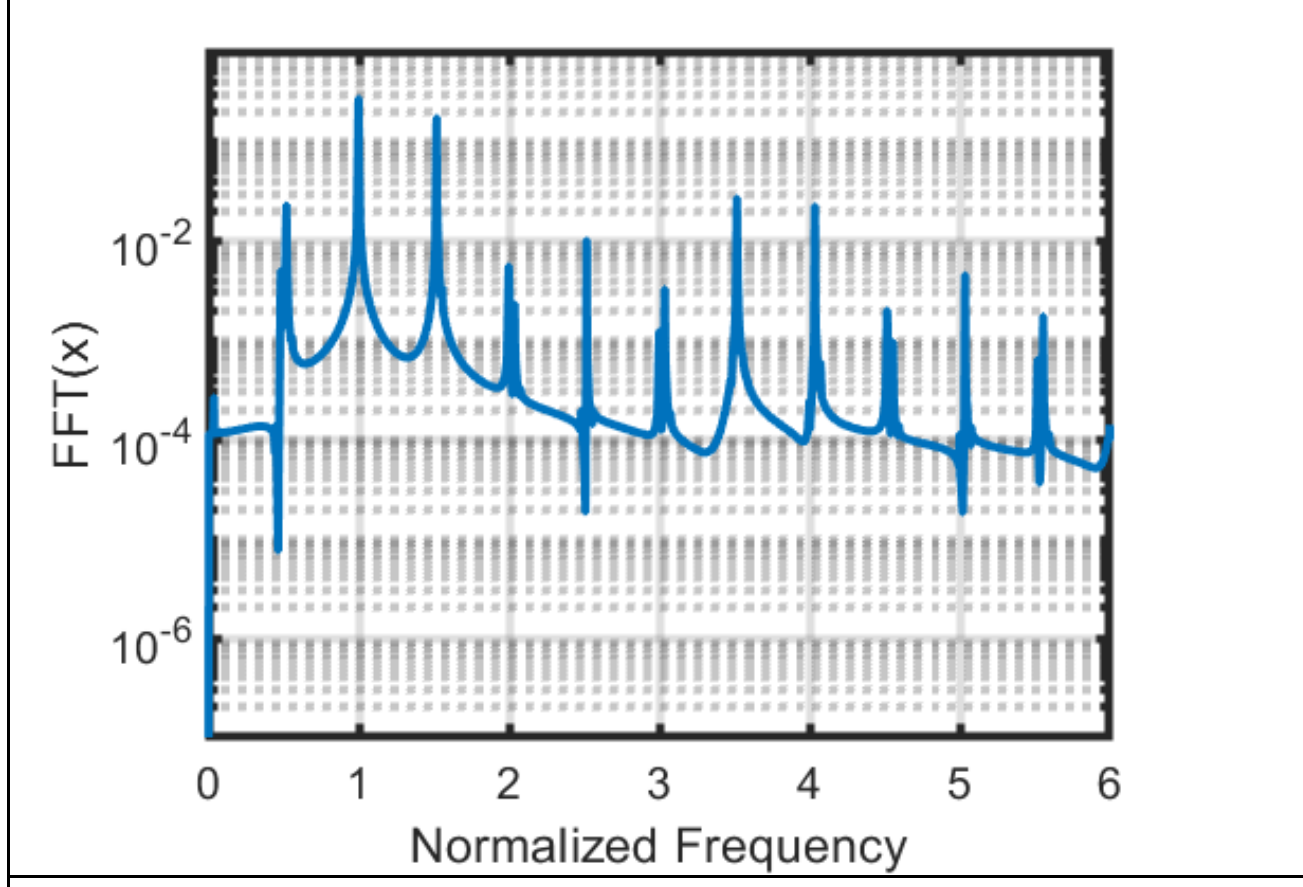

(c)

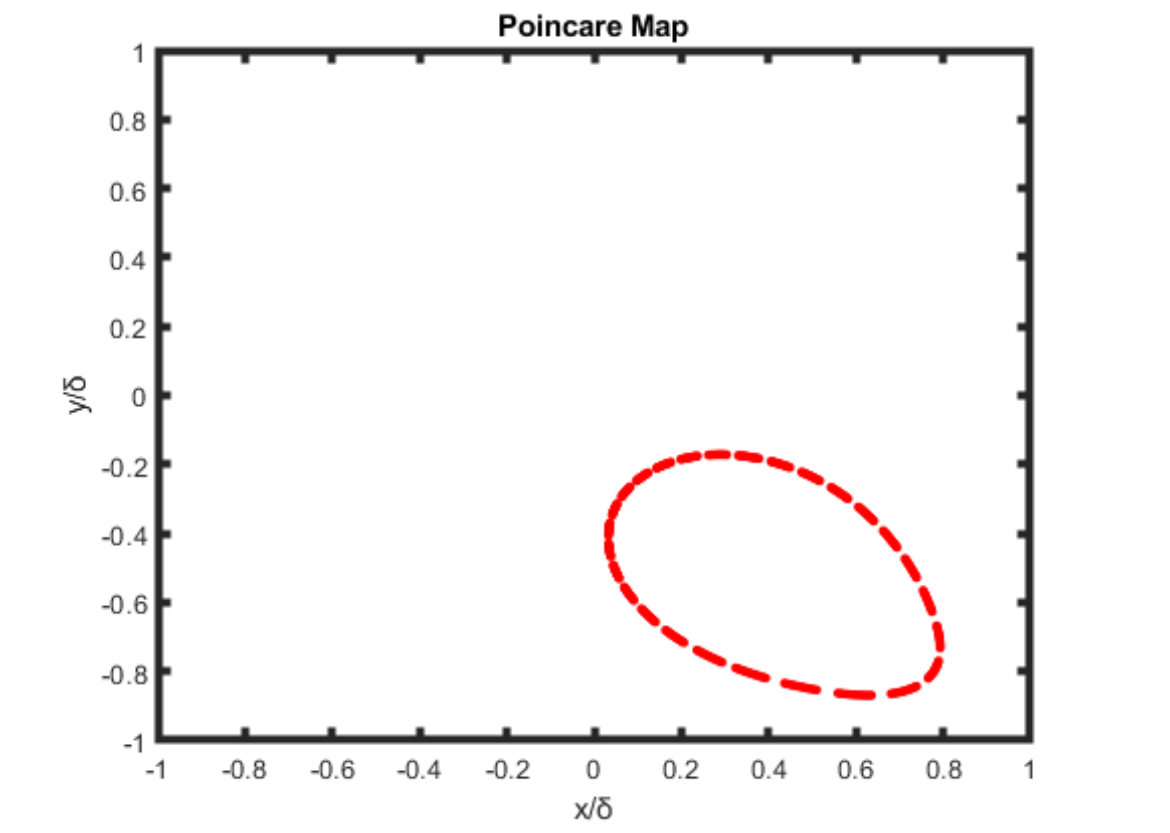

(d)

Fig 6. Rotor Response for case 1 (a) non-dimensional displacement (b) Rotor Orbit (c) FFT (d) Poincare Map

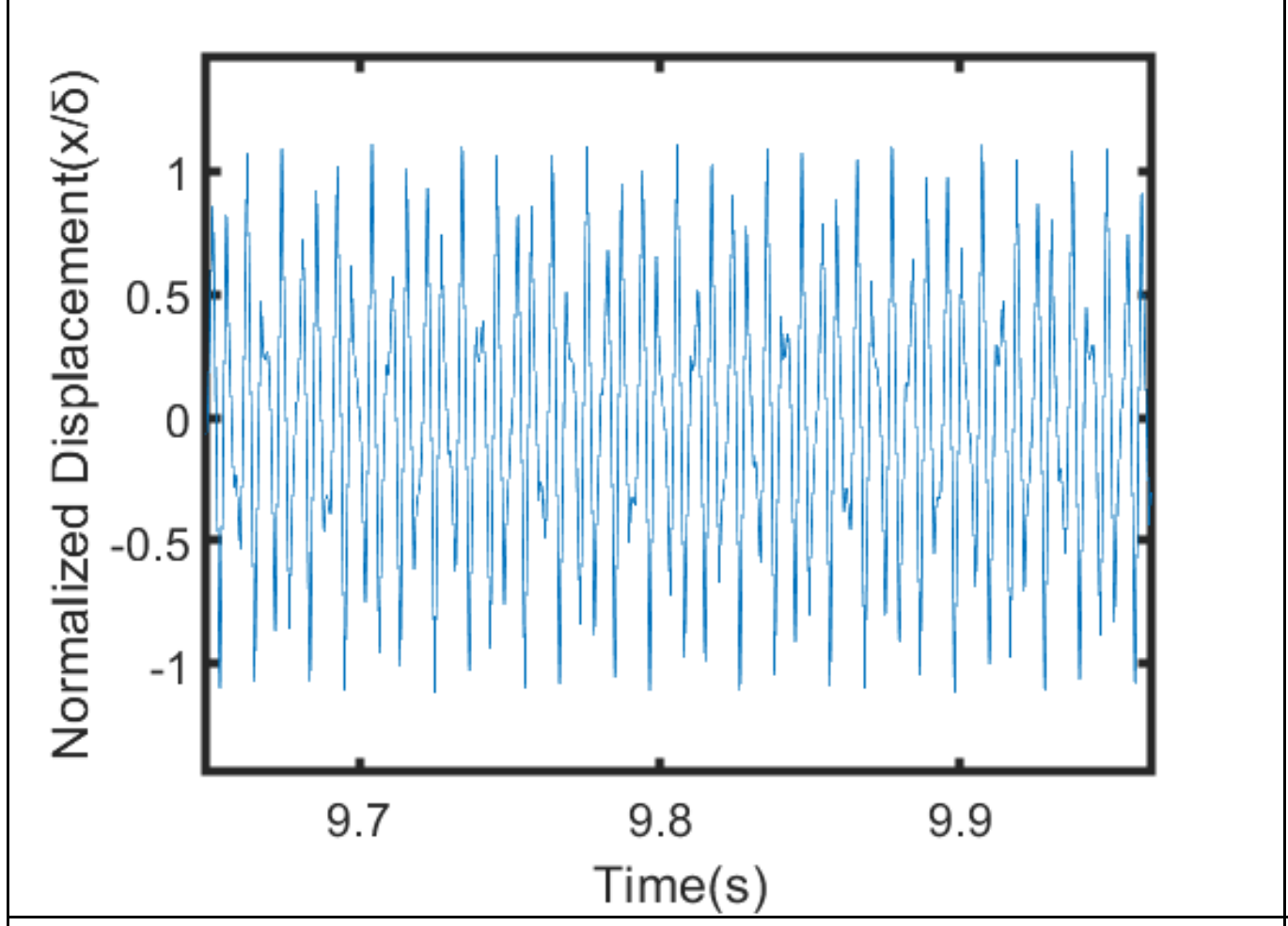

(a)

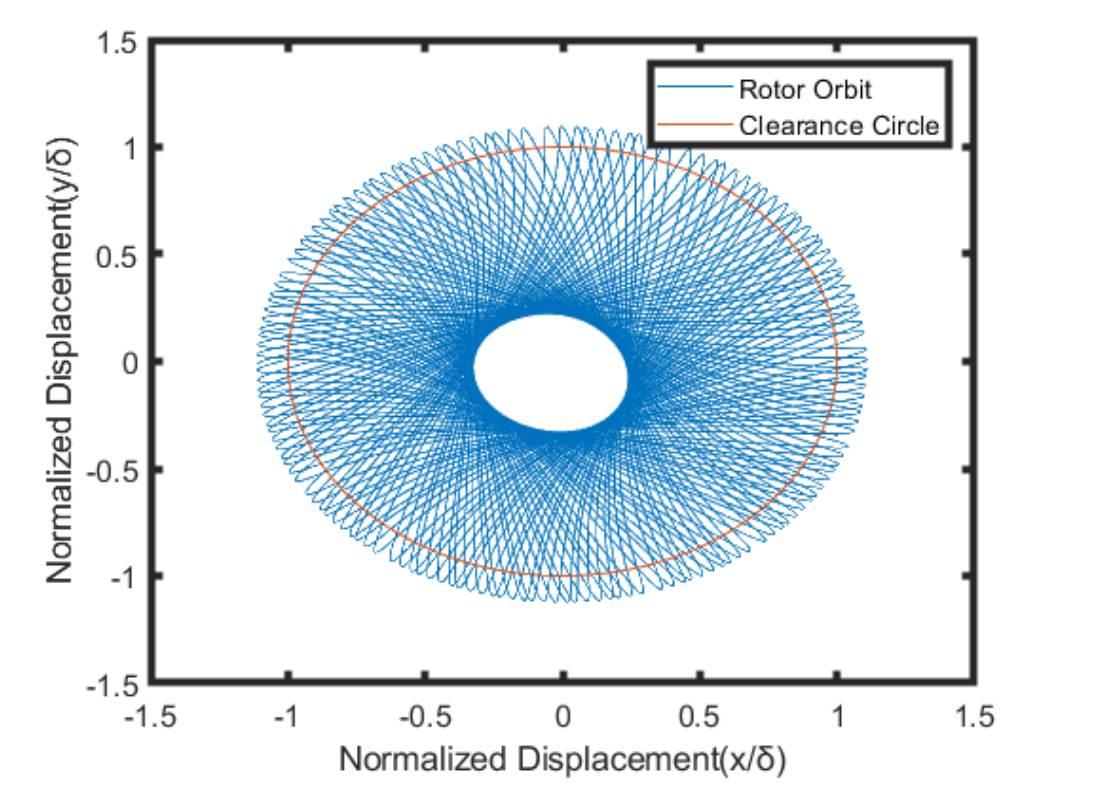

(b)

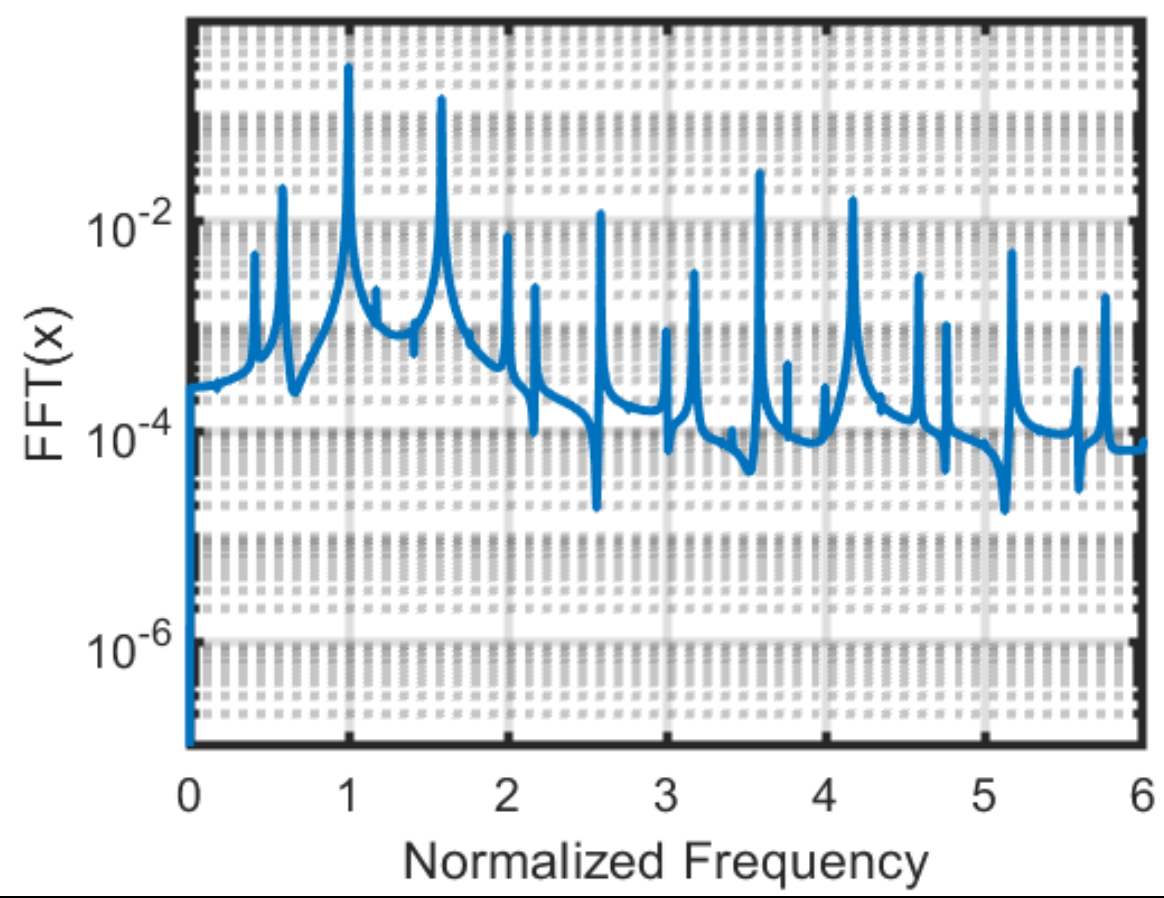

(c)

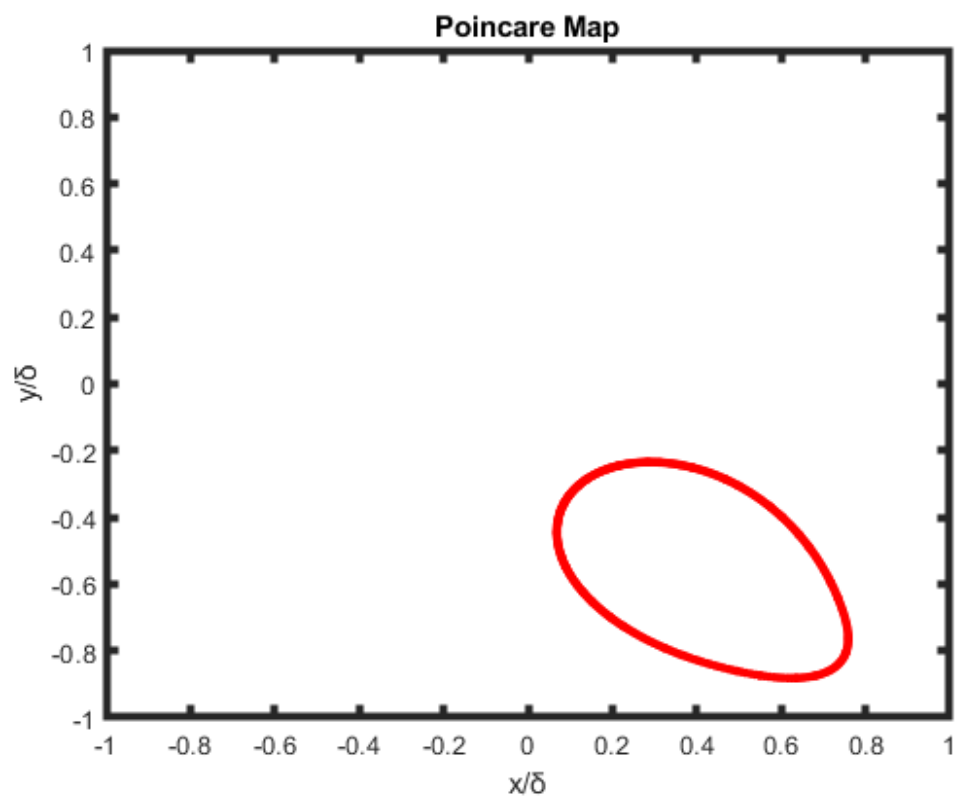

(d)

Fig 7. Rotor Response for case 1 (a) non-dimensional displacement (b) Rotor Orbit (c) FFT (d) Poincare Map

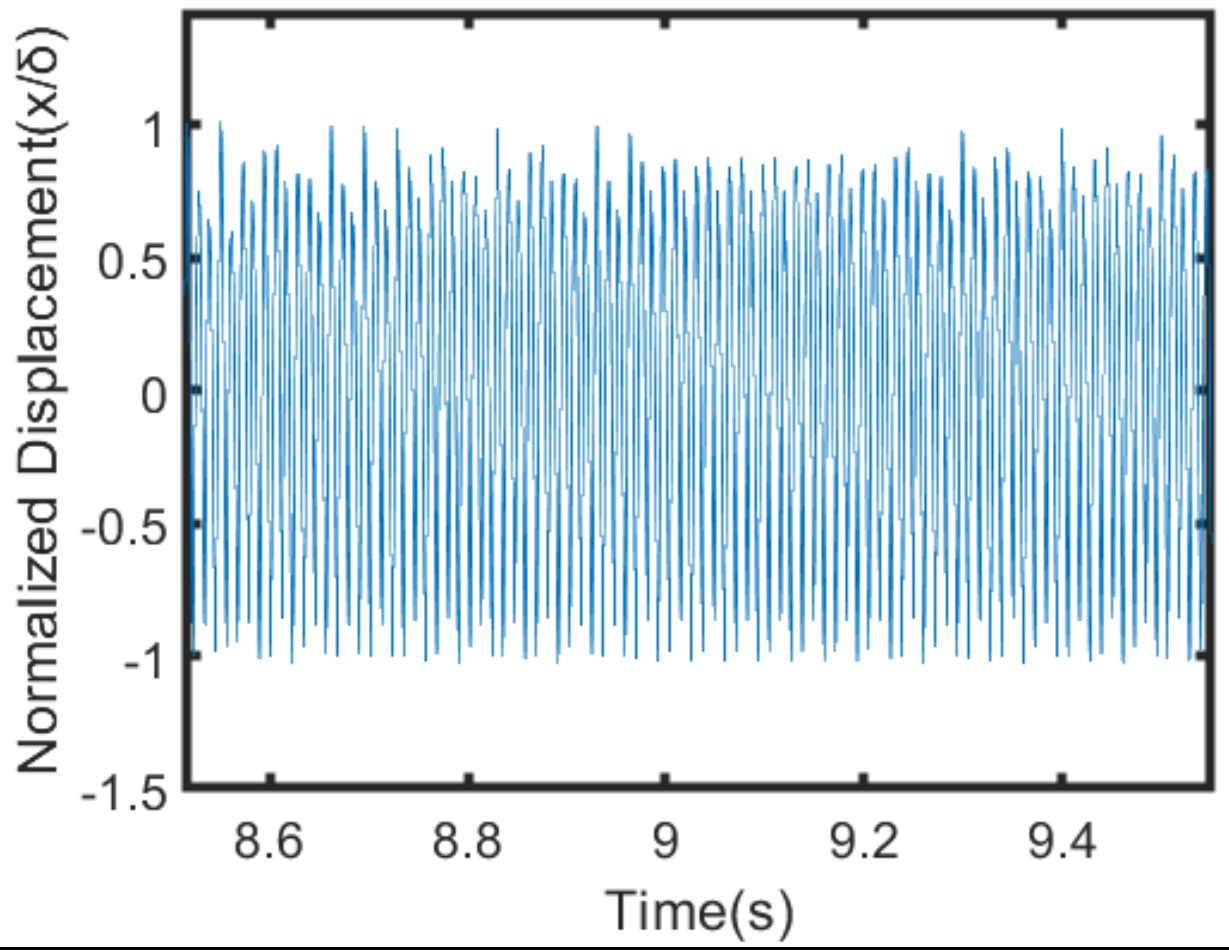

(a)

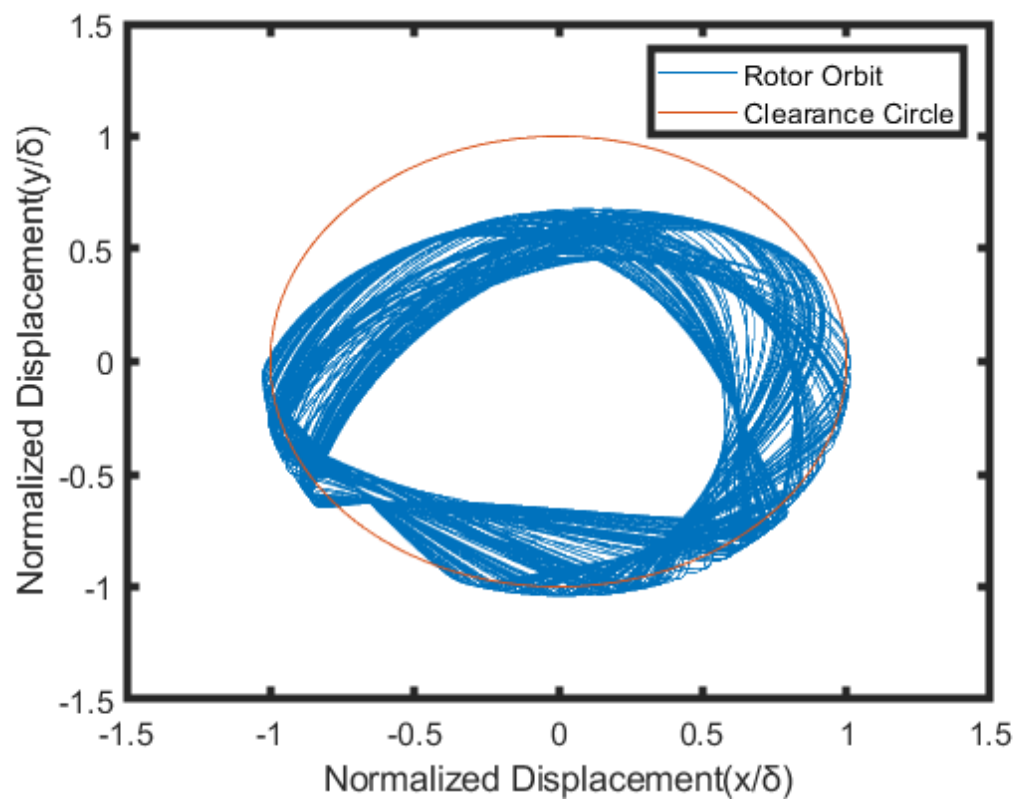

(b)

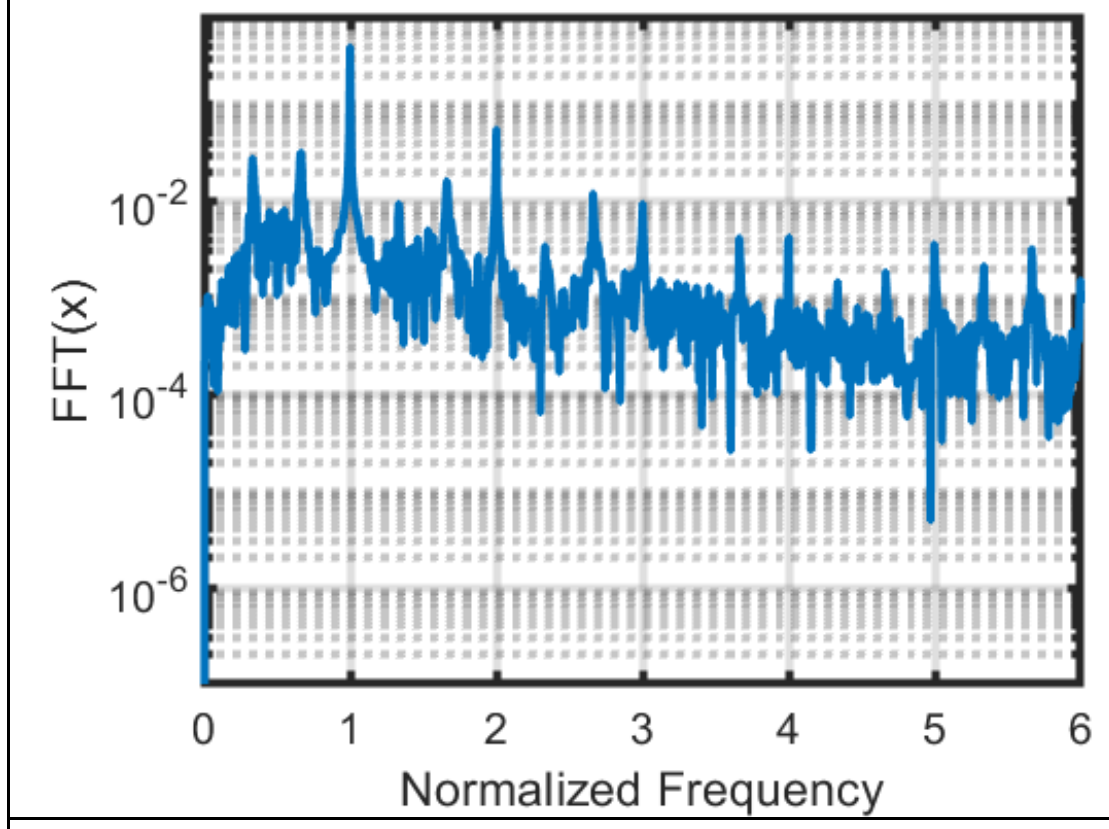

(c)

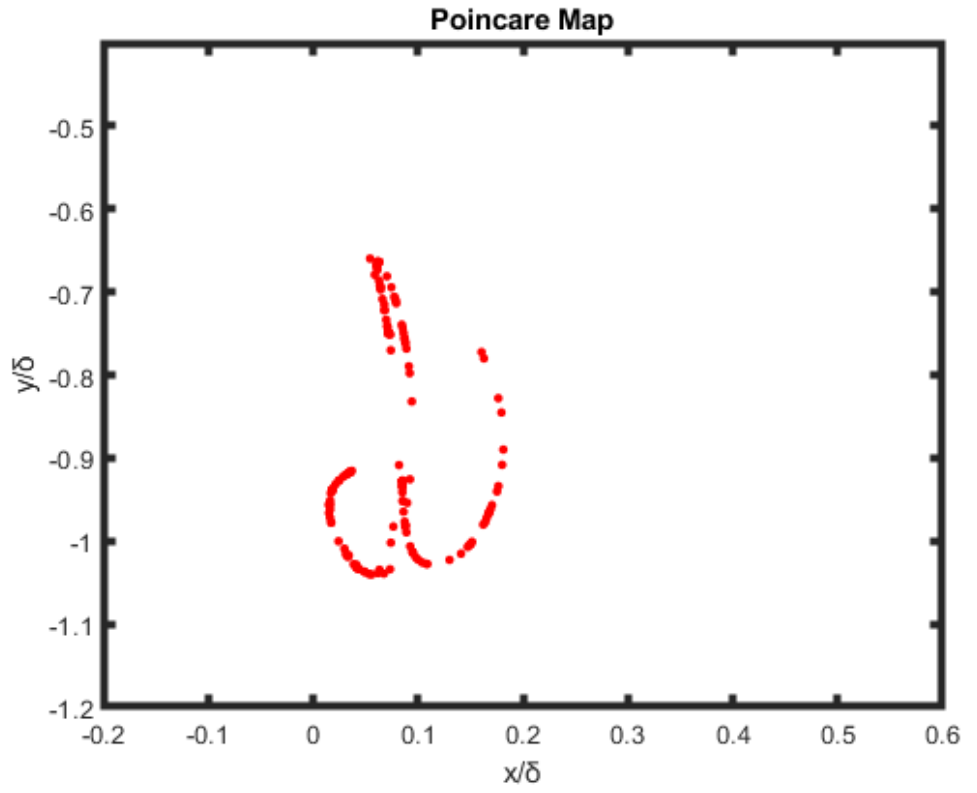

(d)

Fig 8. Rotor Response for case 1 (a) non-dimensional displacement (b) Rotor Orbit (c) FFT (d) Poincare Map

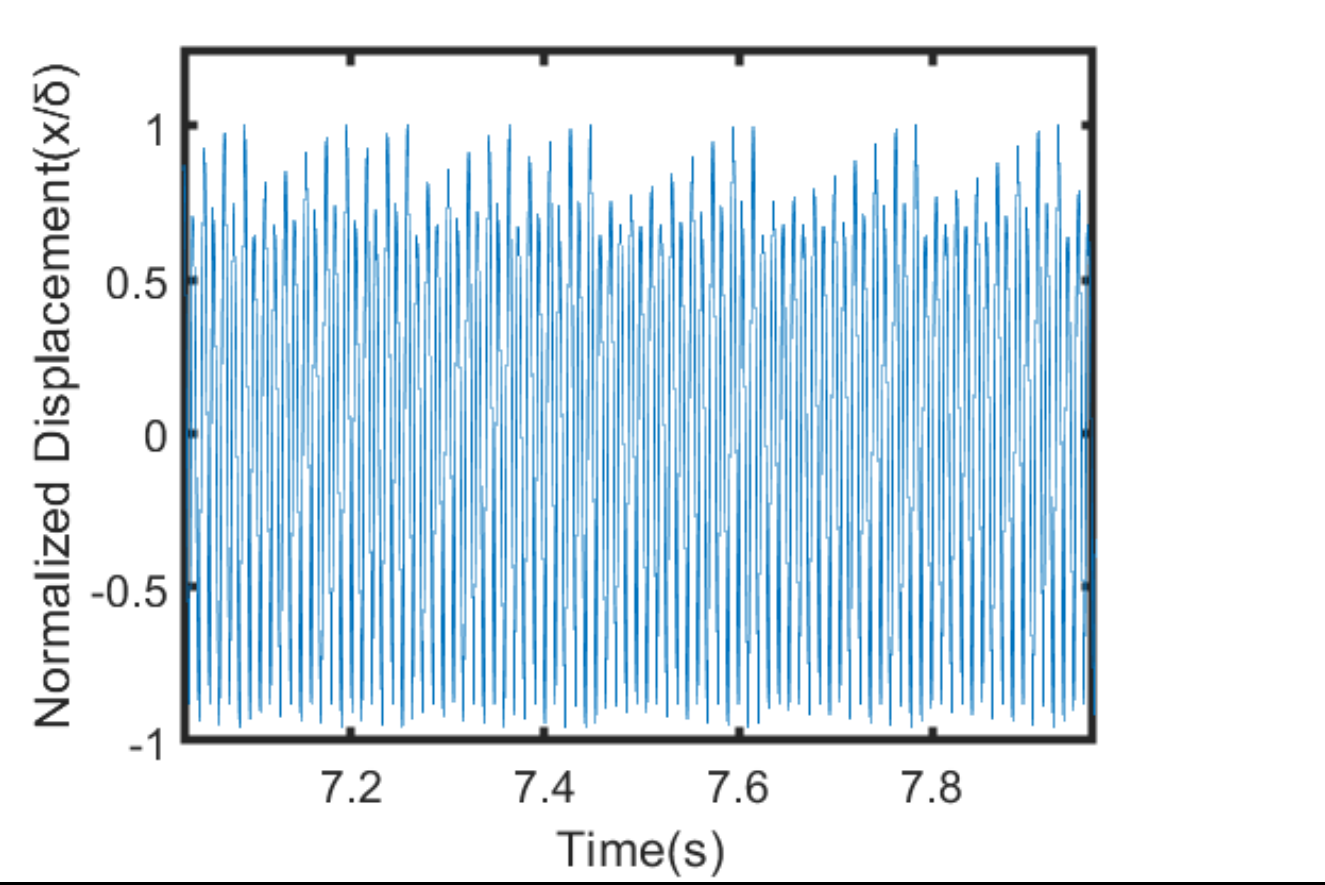

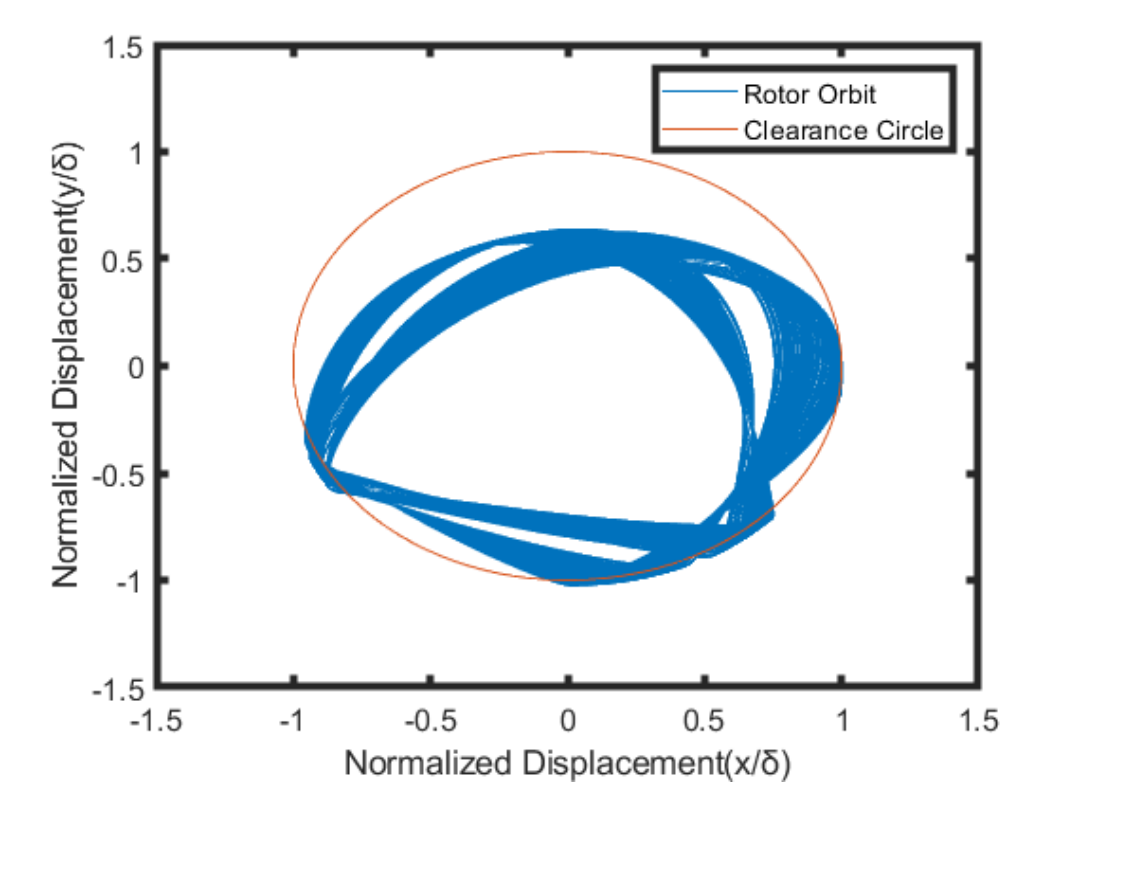

(a) (b)

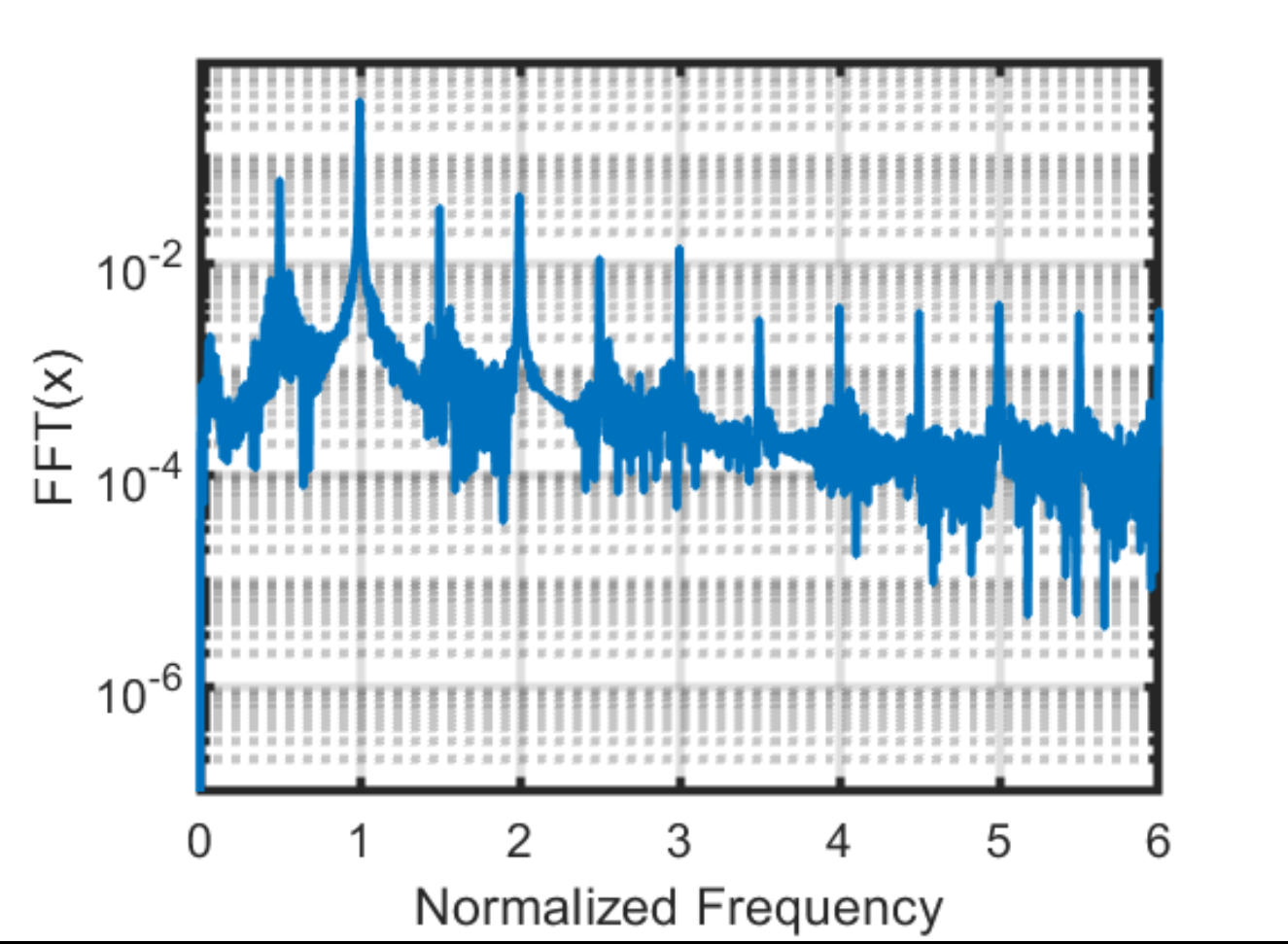

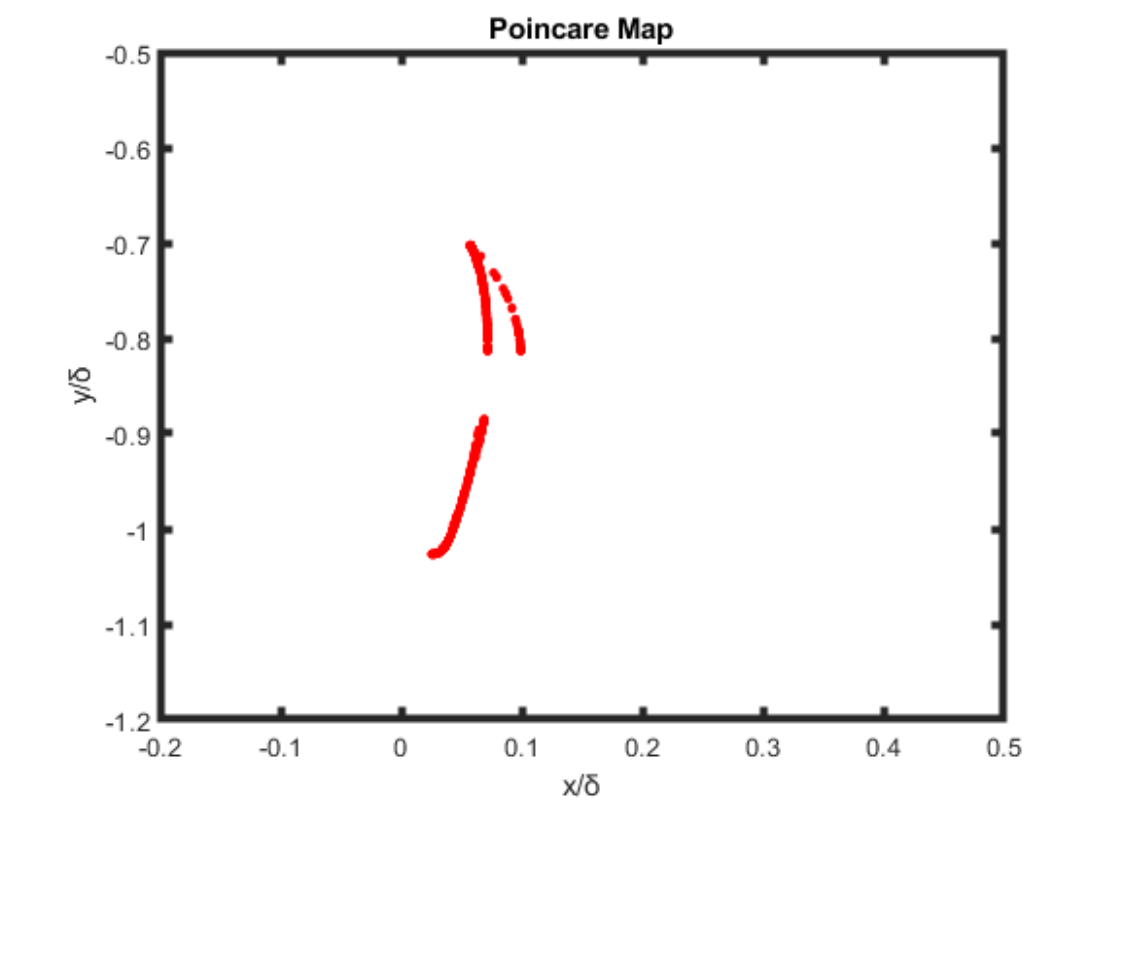

(c) (d)

Fig 9. Rotor Response for case 1 (a) non-dimensional displacement (b) Rotor Orbit (c) FFT (d) Poincare Map

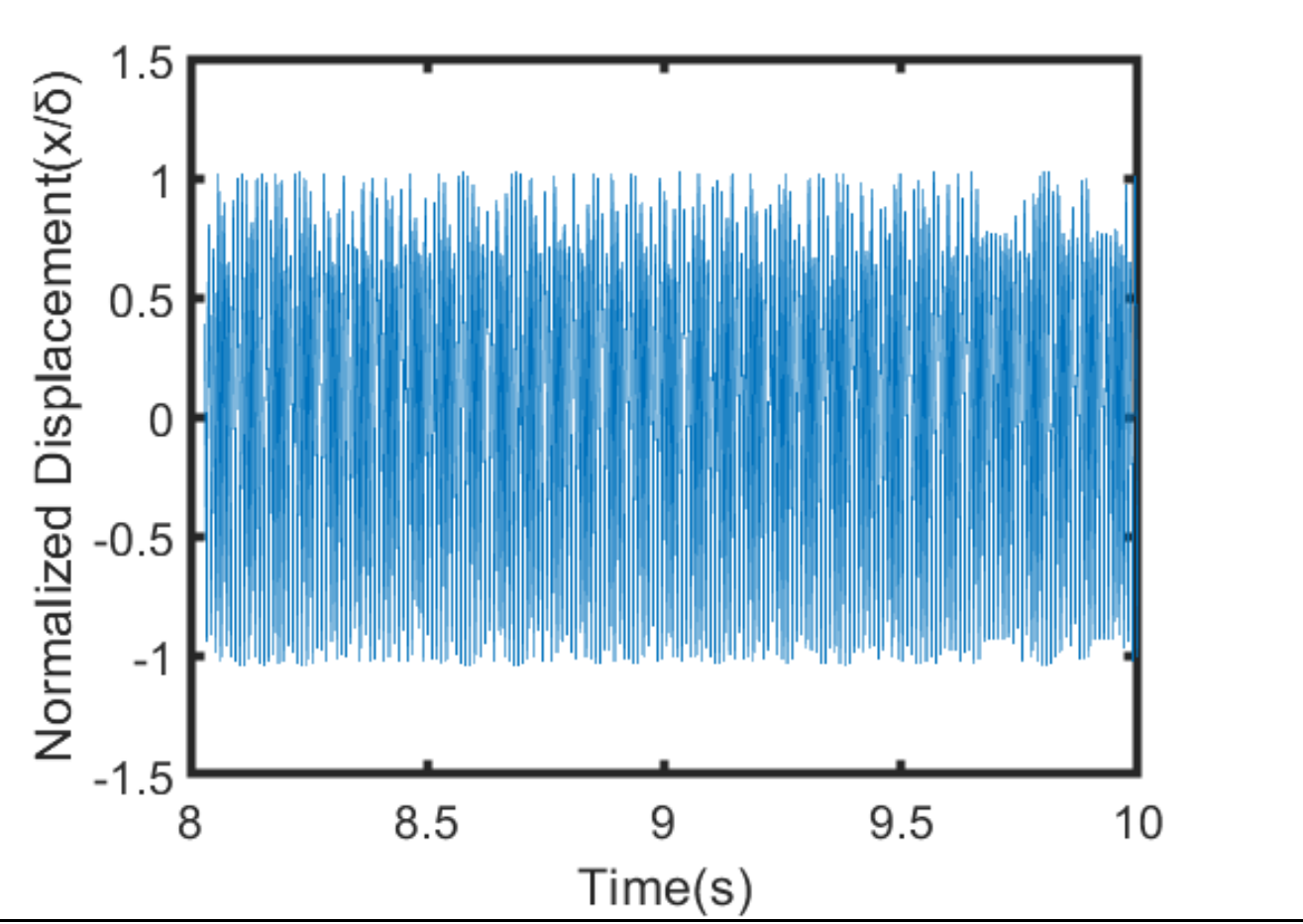

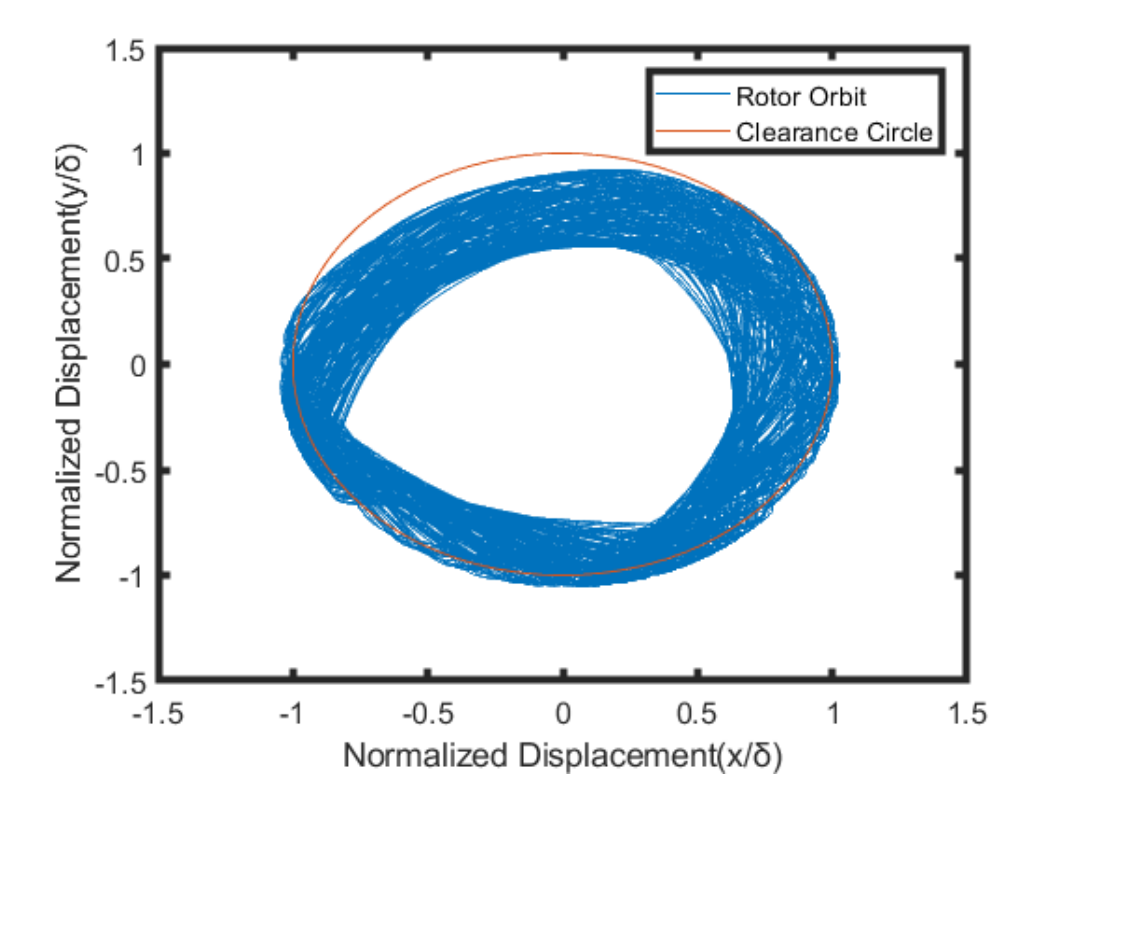

| (a) | (b) |
|---|---|

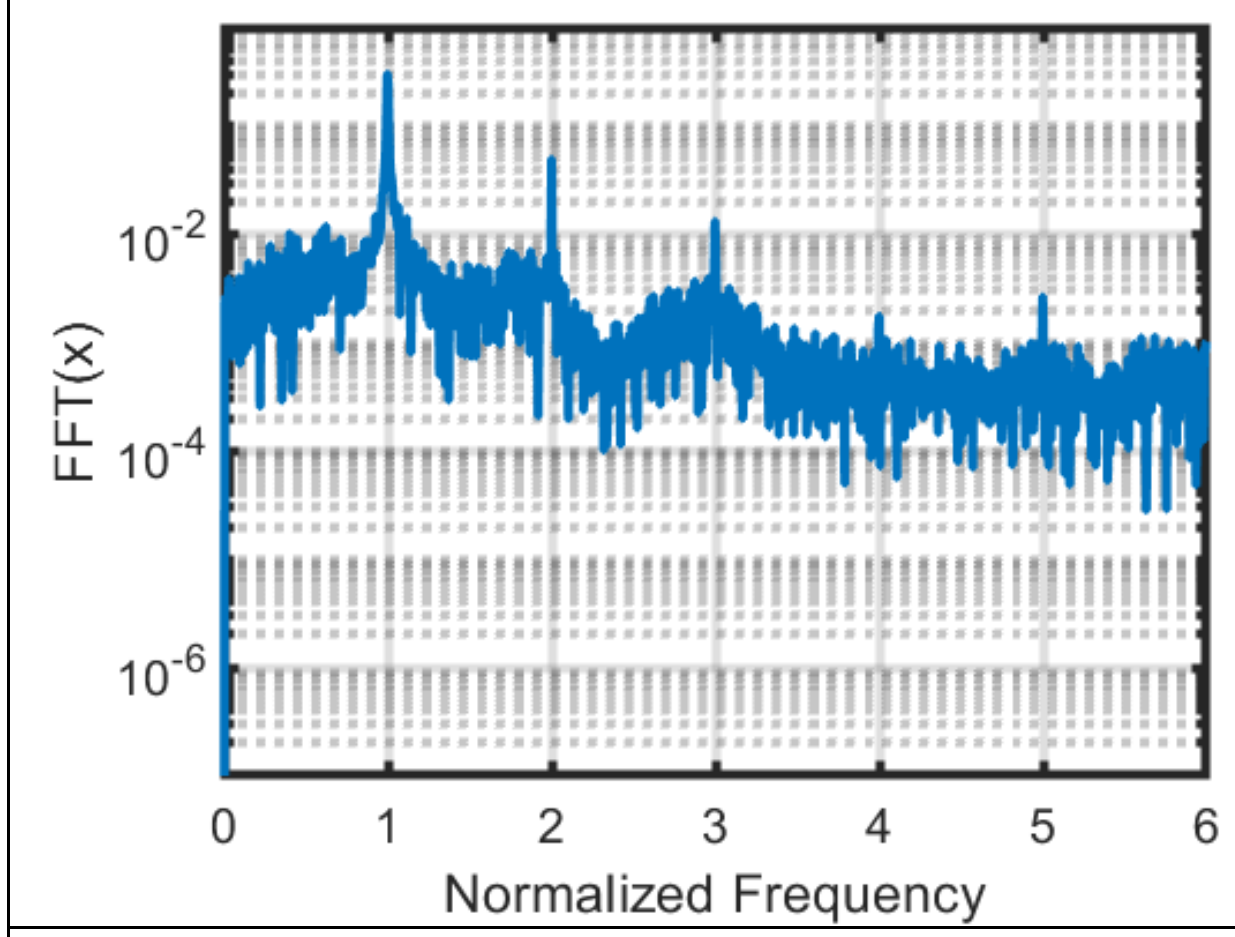

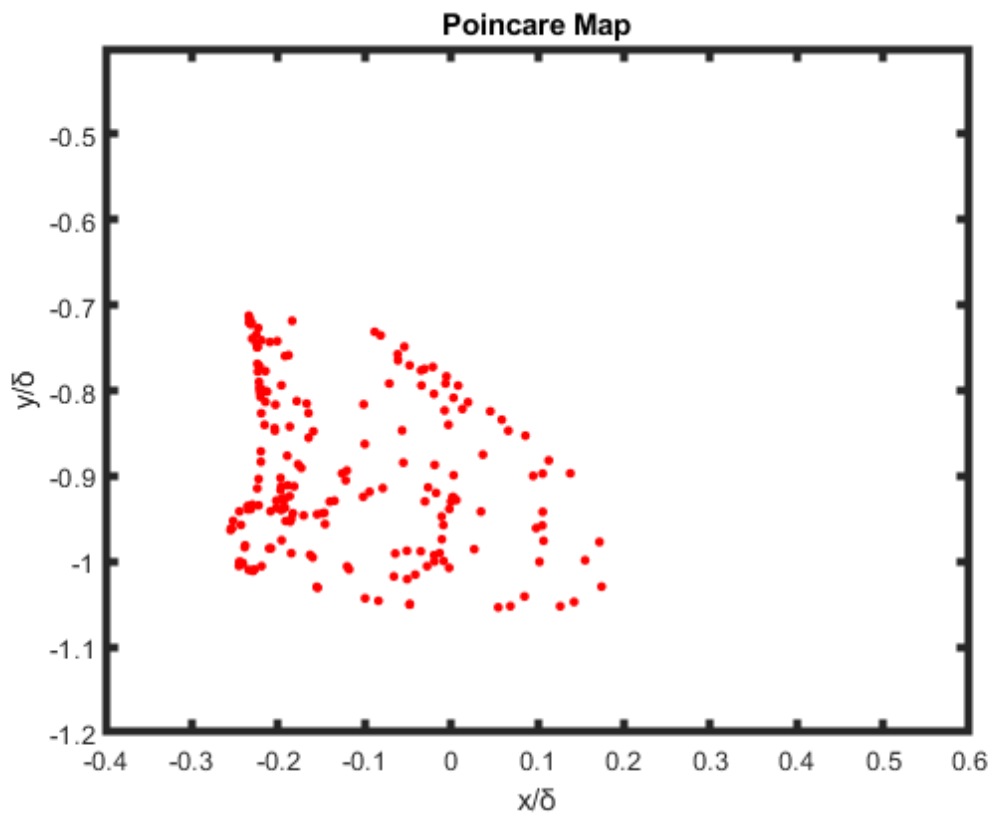

| (c) | (d) |
|---|---|

Fig 10. Rotor Response for case 1 (a) non-dimensional displacement (b) Rotor Orbit (c) FFT (d) Poincare Map

Table 2. K-values from 0-1 test for all the cases of rotor response

| Case | Response | K Value |
|---|---|---|
| 1 | Periodic | 0.05 |
| 2 | Periodic | 0.02 |
| 3 | Periodic | 0.05 |
| 4 | Quasi-Periodic | 0.05 |
| 5 | Quasi-Periodic | 0.039 |
| 6 | Quasi-Periodic | 0.035 |
| 7 | Chaotic | 0.986 |
| 8 | Chaotic | 0.969 |
| 9 | Chaotic | 0.997 |

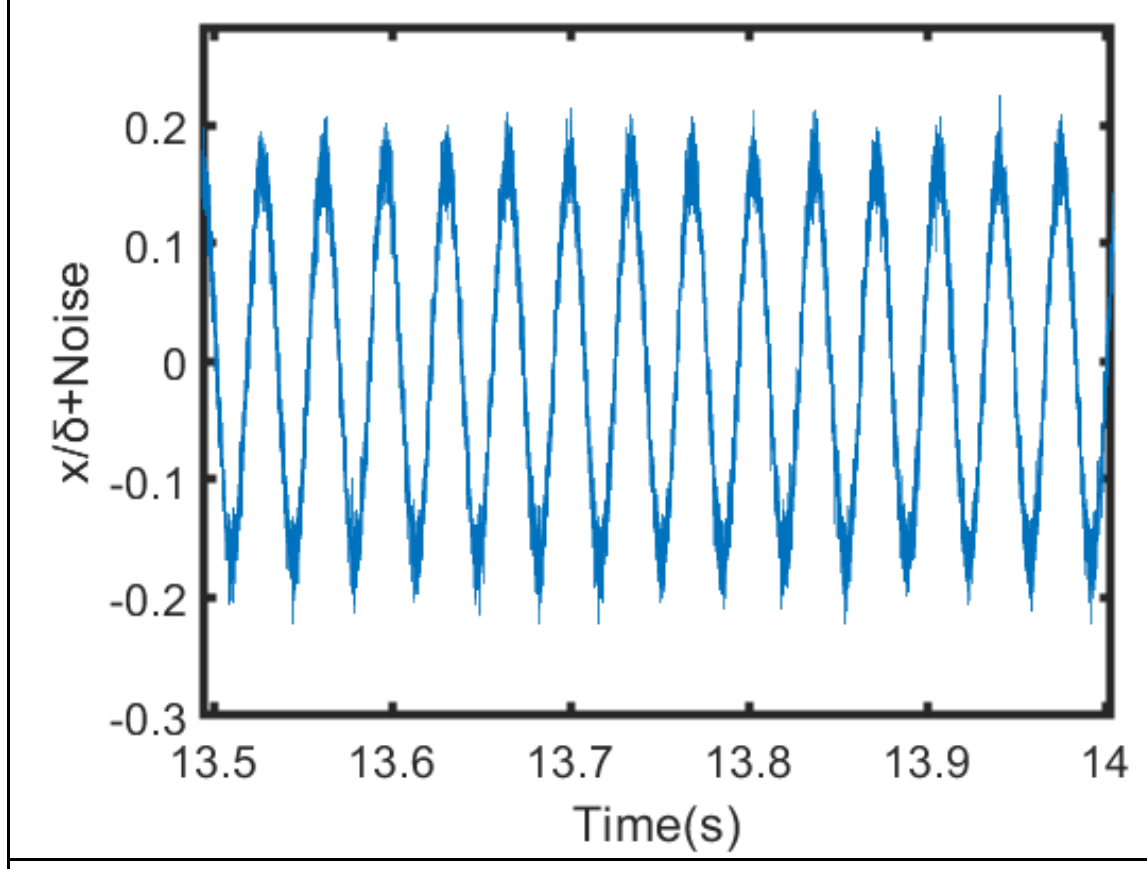

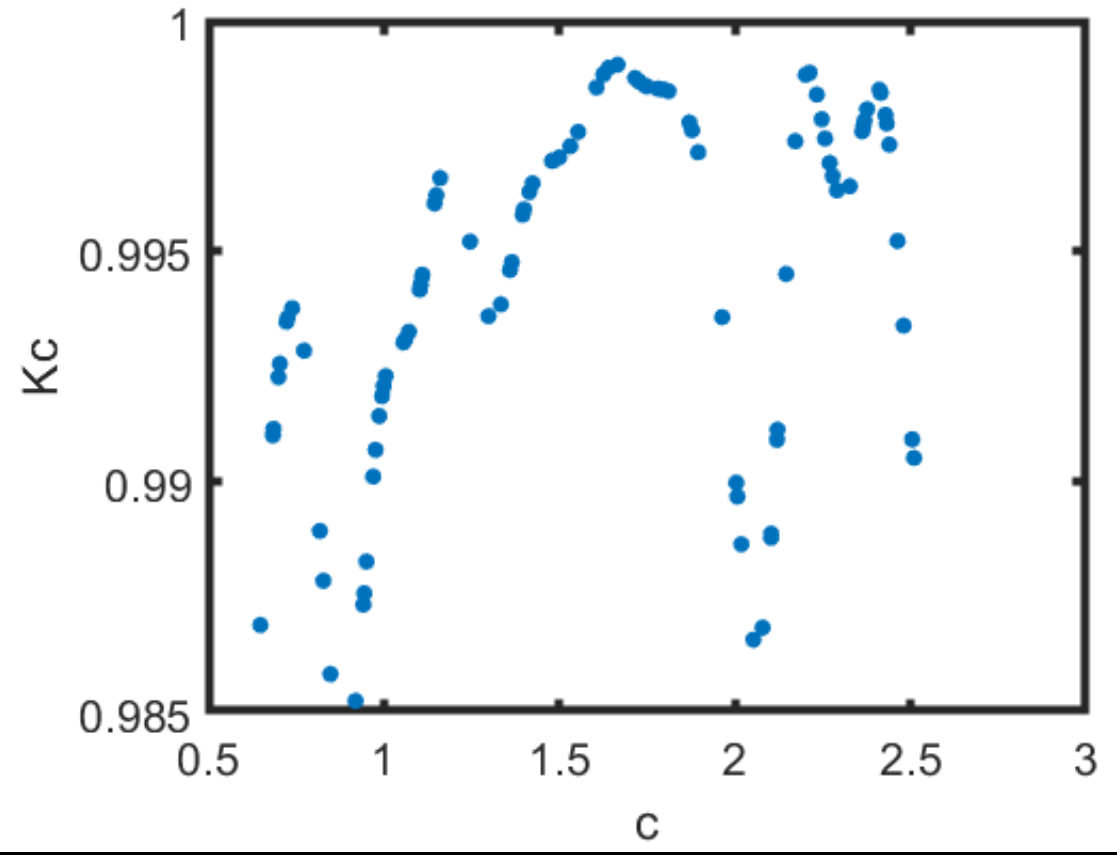

| (a) | (b) |
|---|---|

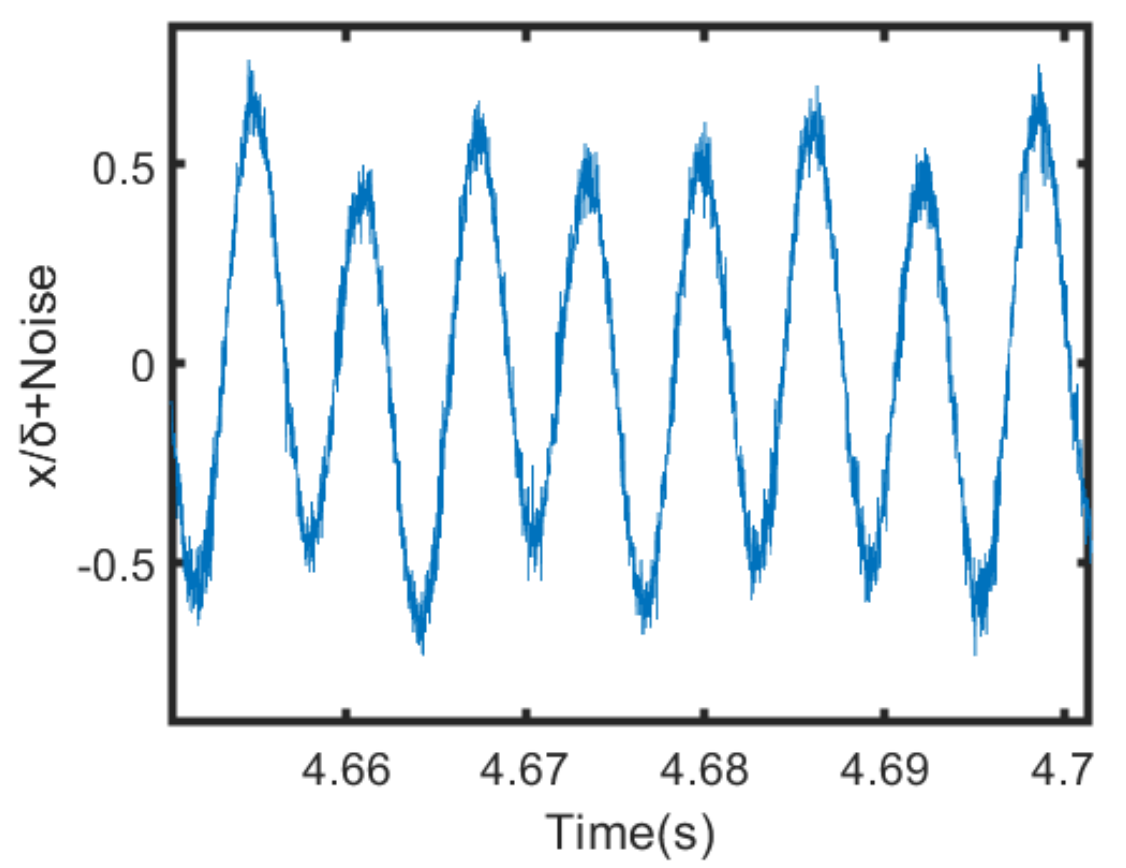

(c)

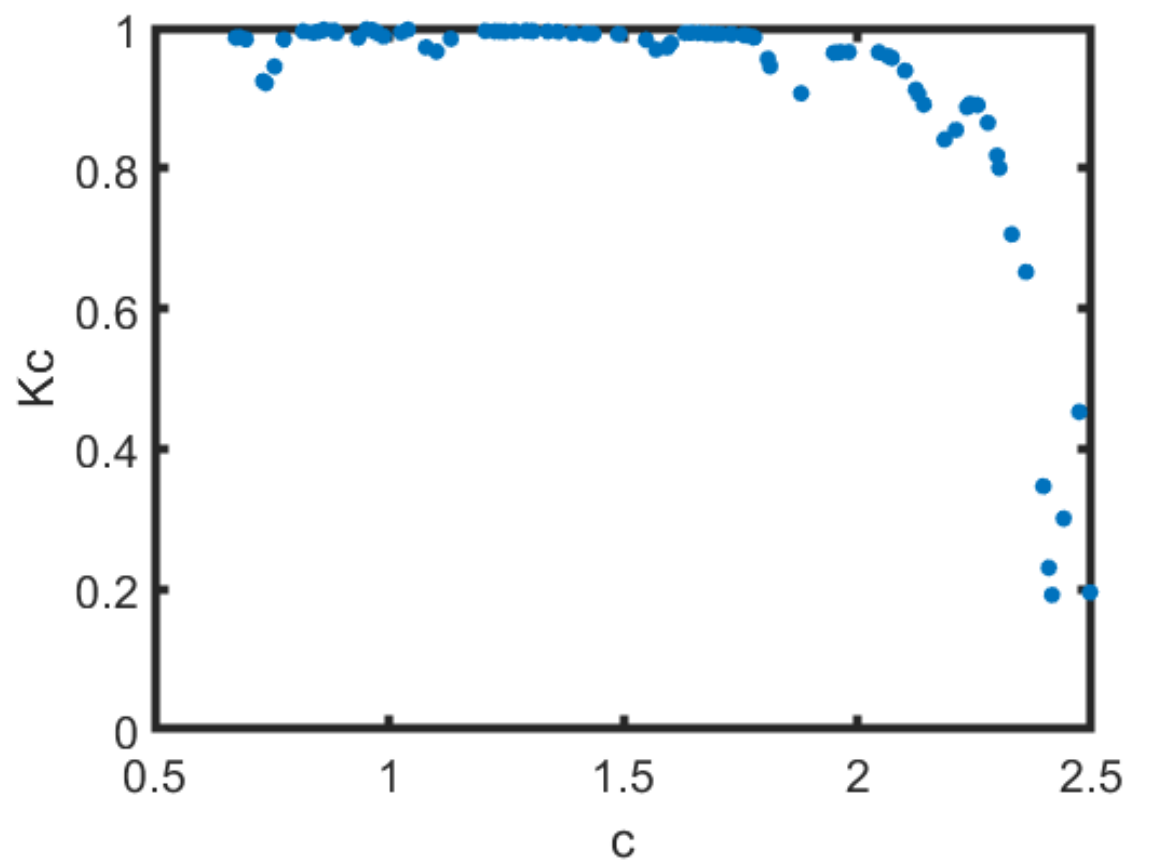

(d)

Fig 11. Signal mixed with noise ( SNR value 50) (a) Case 1 signal with WGN (b) K-values for case 1 signal with WGN (c) Case 4 signal with WGN (d) K-values for case 4 signal with WGN

Table 3. 0-1 test results on Signal mixed with Noise

| Noisy Signal Case | Primary Signal (From Table 1) | Noise Type | K Value | 0-1 Test Result(Signal Type) | Actual Signal Type |
|---|---|---|---|---|---|
| C1 | Case 1 | White Gaussian Noise | 0.9990 | Chaotic | Periodic+Noise |
| C2 | Case 2 | White Gaussian Noise | 0.9976 | Chaotic | Periodic+Noise |
| C3 | Case 3 | White Gaussian Noise | 0.9969 | Chaotic | Periodic+Noise |
| C4 | Case 4 | White Gaussian Noise | 0.9987 | Chaotic | Quasi-Periodic+Noise |
| C5 | Case 5 | White Gaussian Noise | 0.9980 | Chaotic | Quasi-Periodic+Noise |
| C6 | Case 6 | White Gaussian Noise | 0.9985 | Chaotic | Quasi-Periodic+Noise |

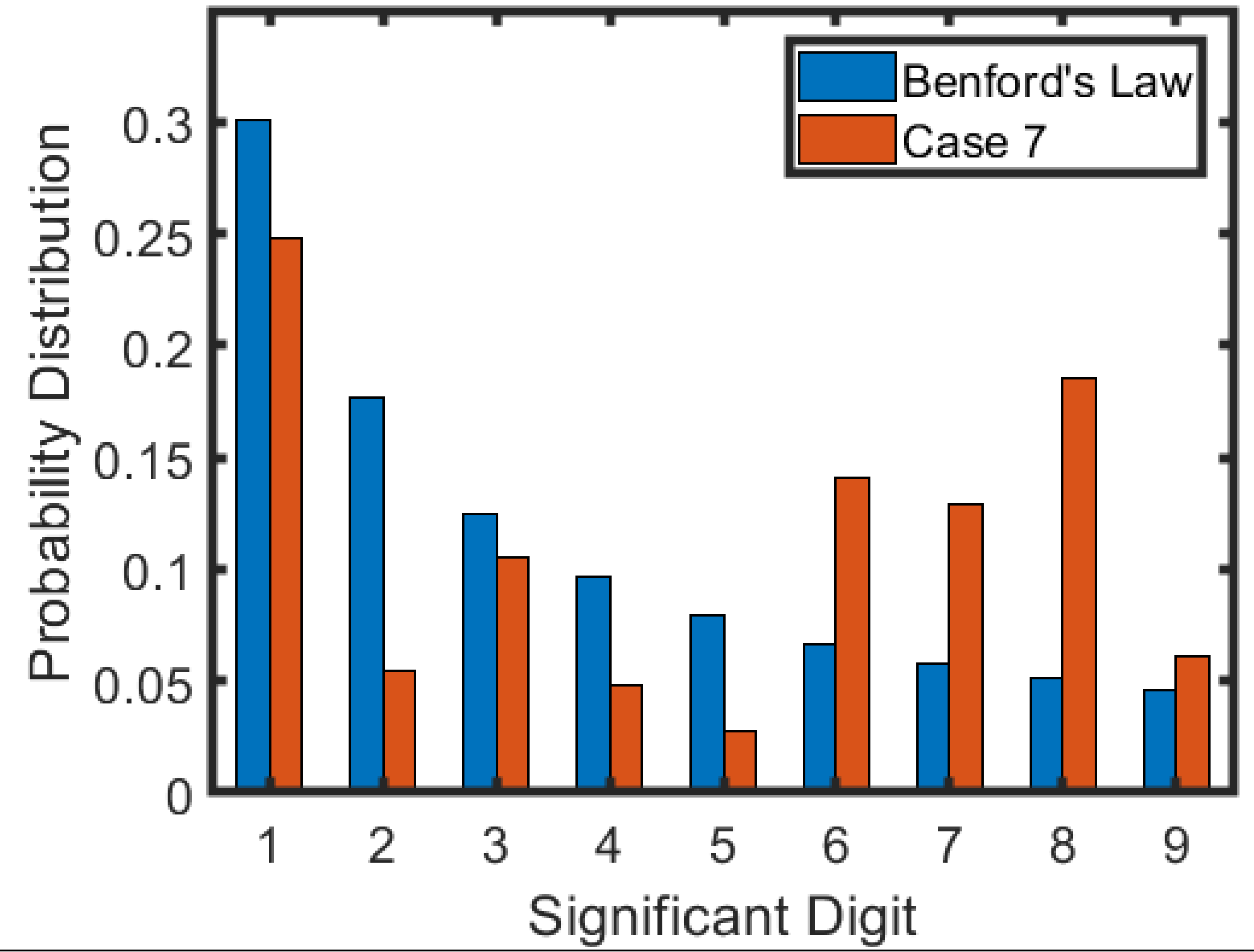

Figure 12(a) Distribution of significant Digit in White Gaussian Noise compared with Benford’s Law

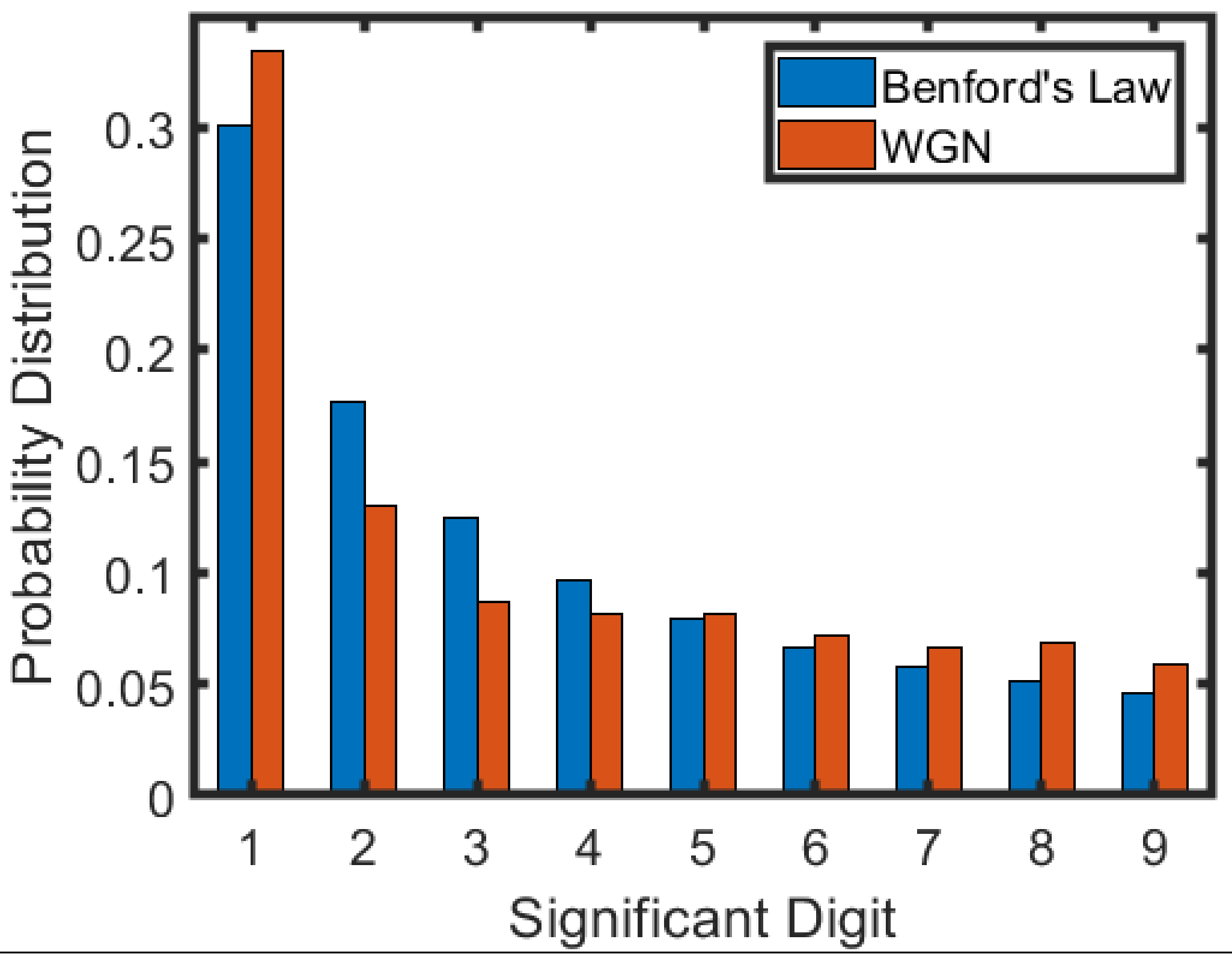

Figure 12(b) Distribution of significant Digit in White Gaussian Noise compared with Benford's Law

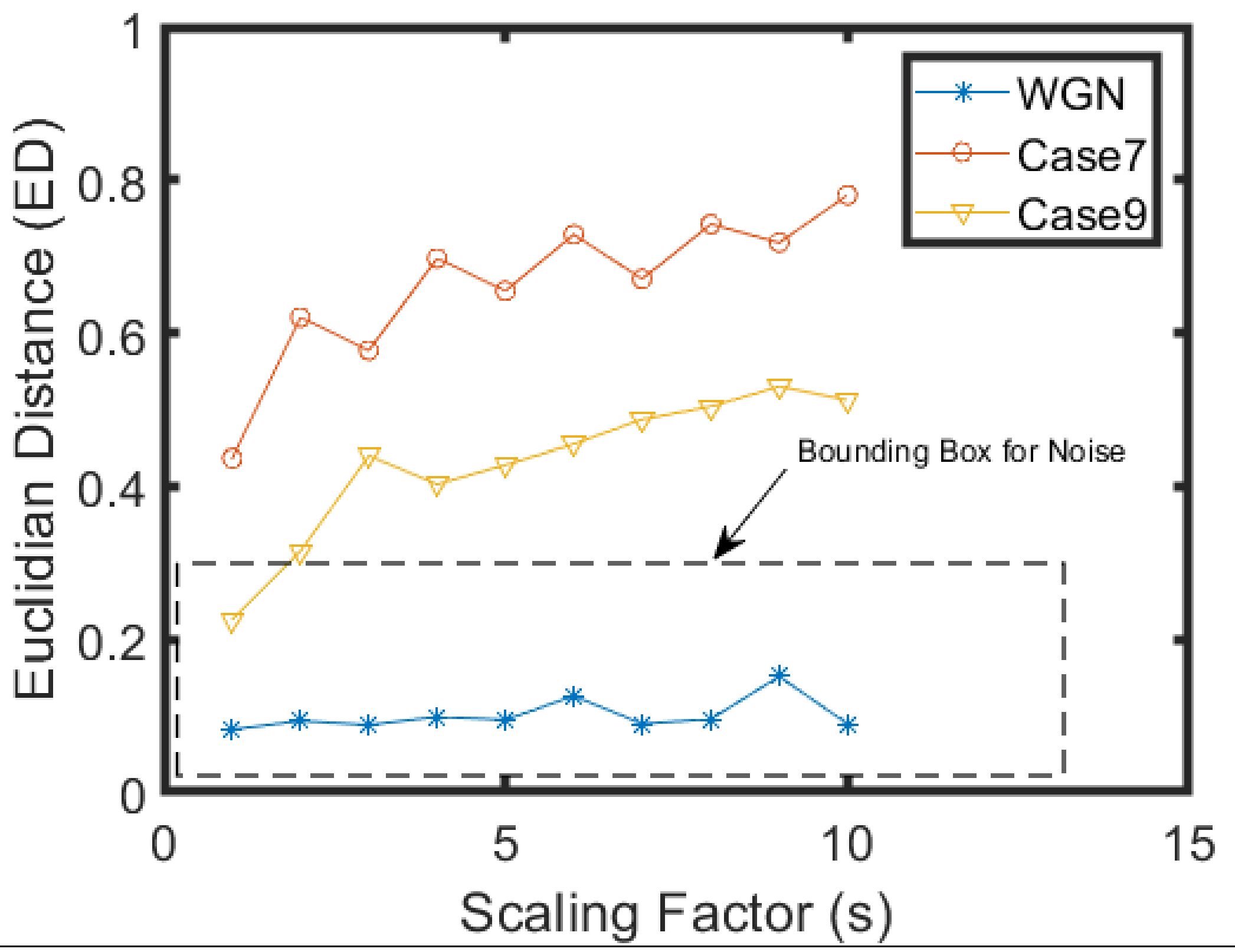

Fig 13. ED variation with scale for chaotic response compared with White Gaussian Noise

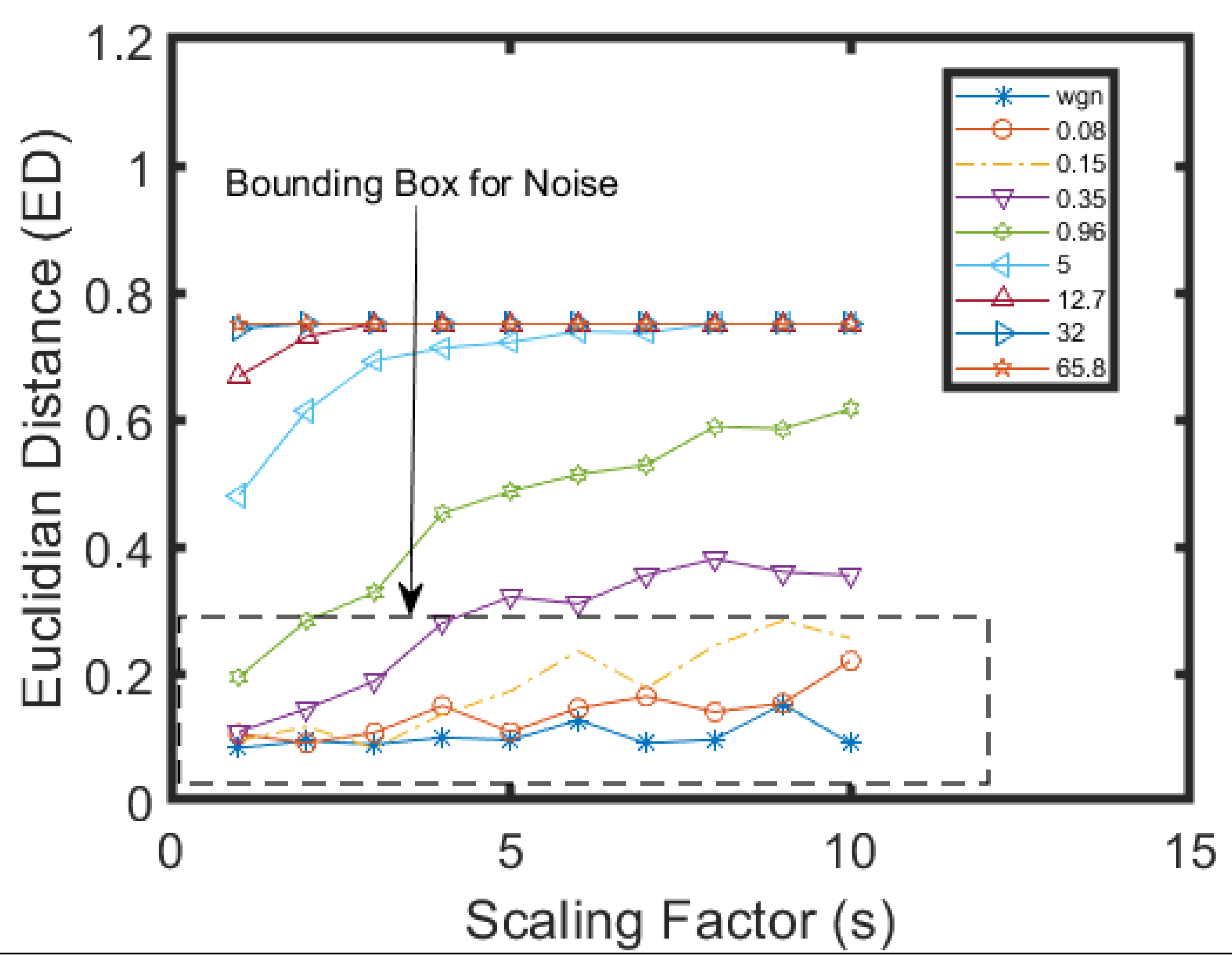

Fig 14. ED variation with scale for periodic response (case 1) mixed with different levels of WGN

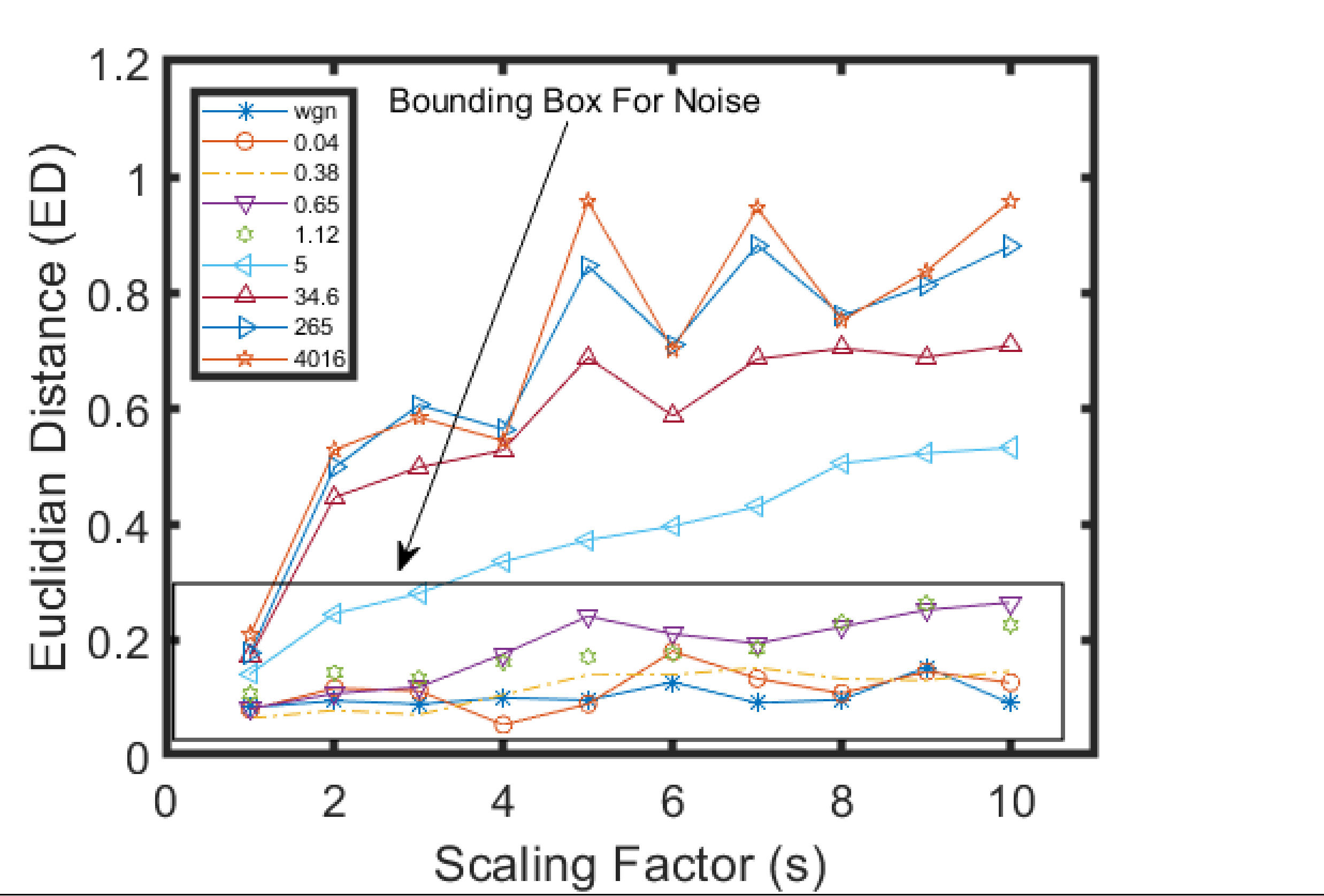

Fig 15. ED variation with scale for quasi-periodic response (case 6) mixed with different levels of WGN

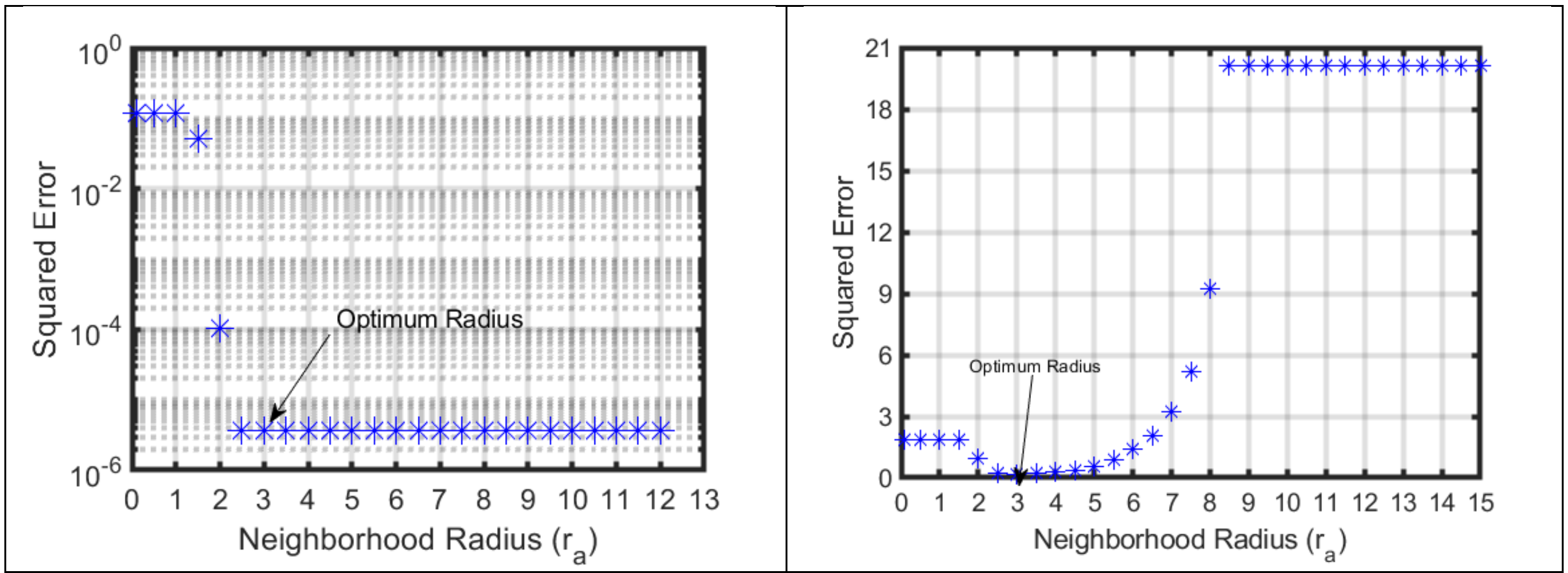

(a) (b)

Fig 16. Optimum Neighborhood radius parameter ($r_a$) for (a) periodic case 1 (b) Quasi-periodic case 6, both with SNR value 30.

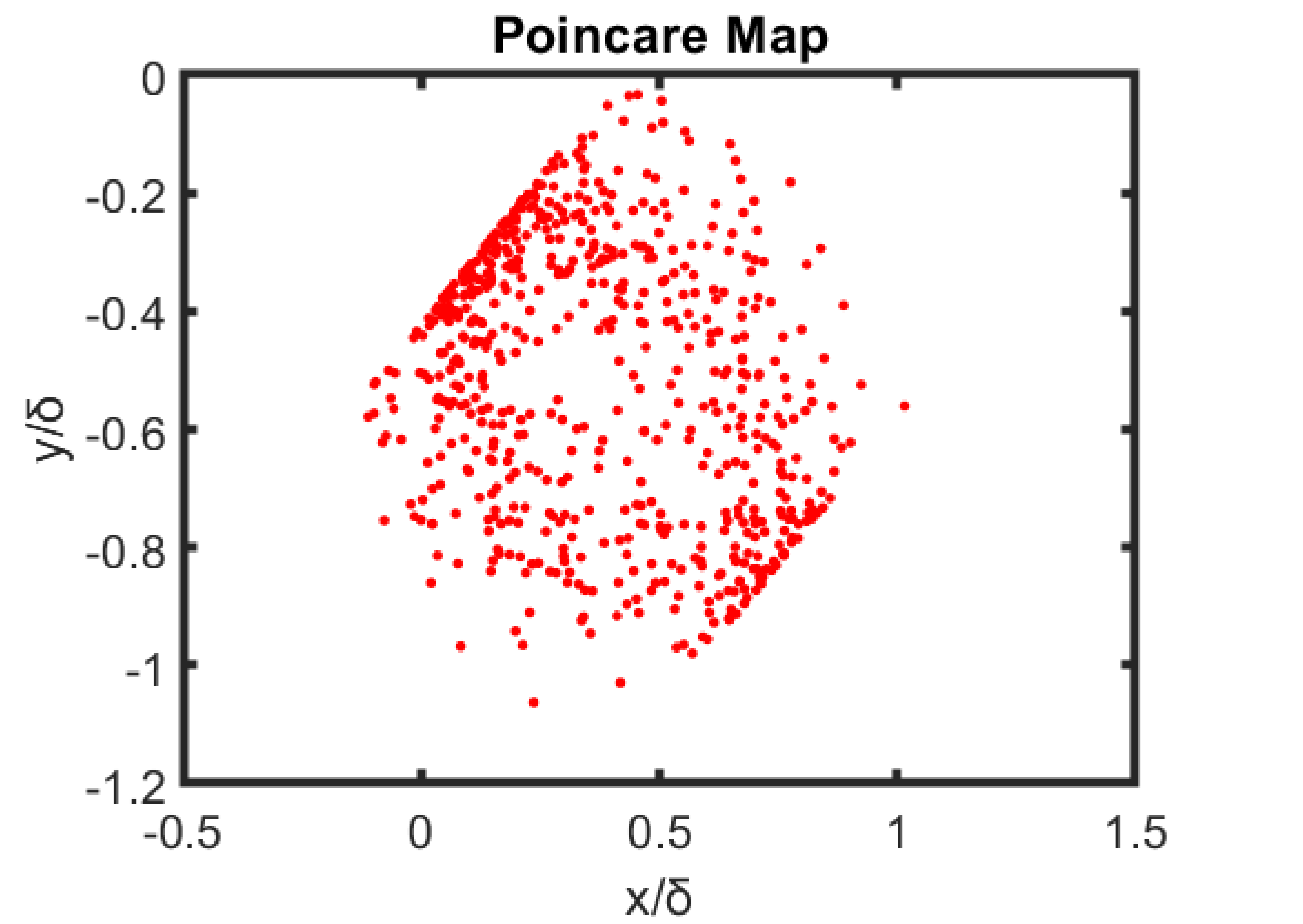

Figure 17 (a). Poincare Map of Quasi-periodic Vibration (case 6) with SNR 45

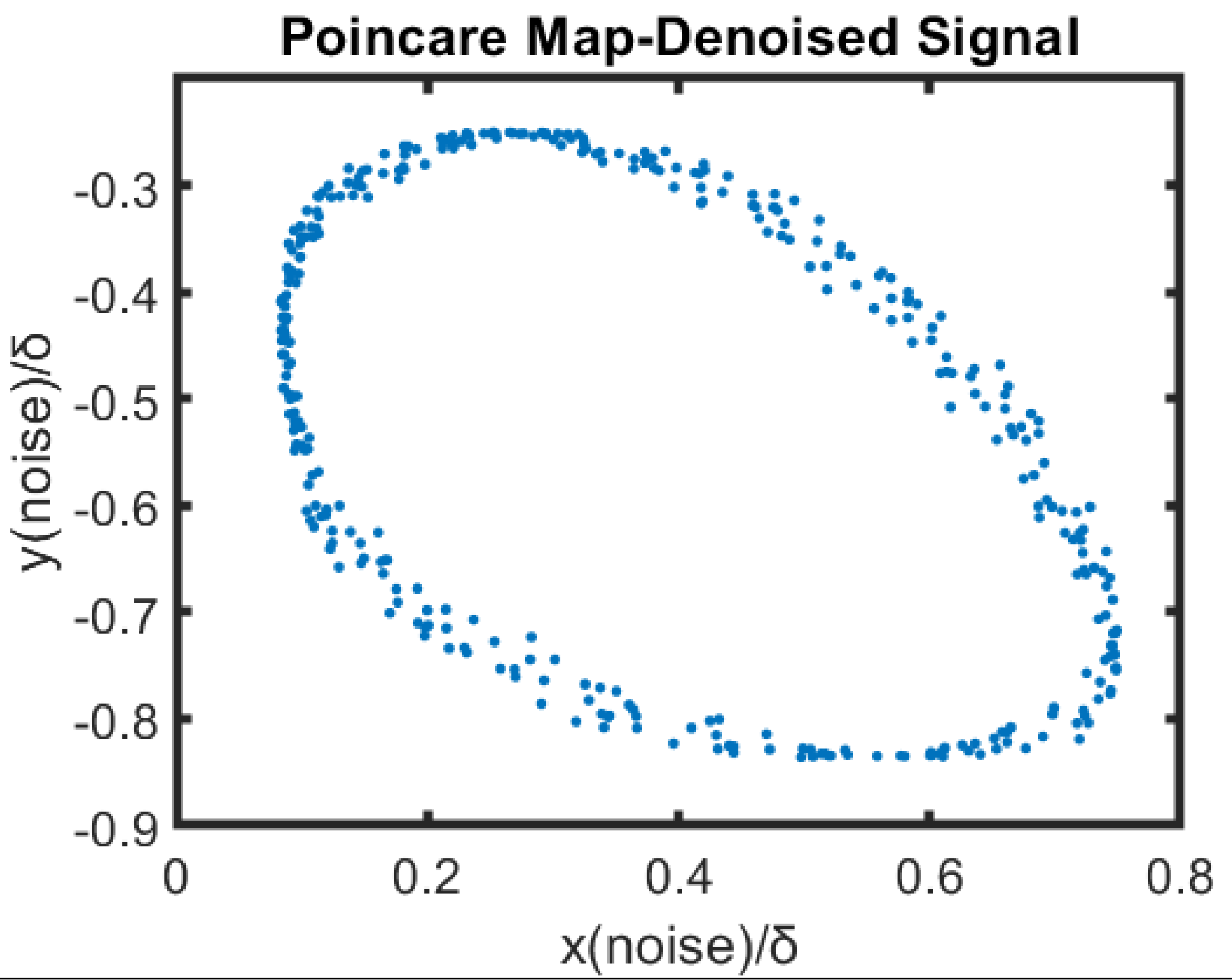

Figure 17 (b). Poincare Map of Quasi-periodic Vibration (case 6) with SNR 45 after noise reduction

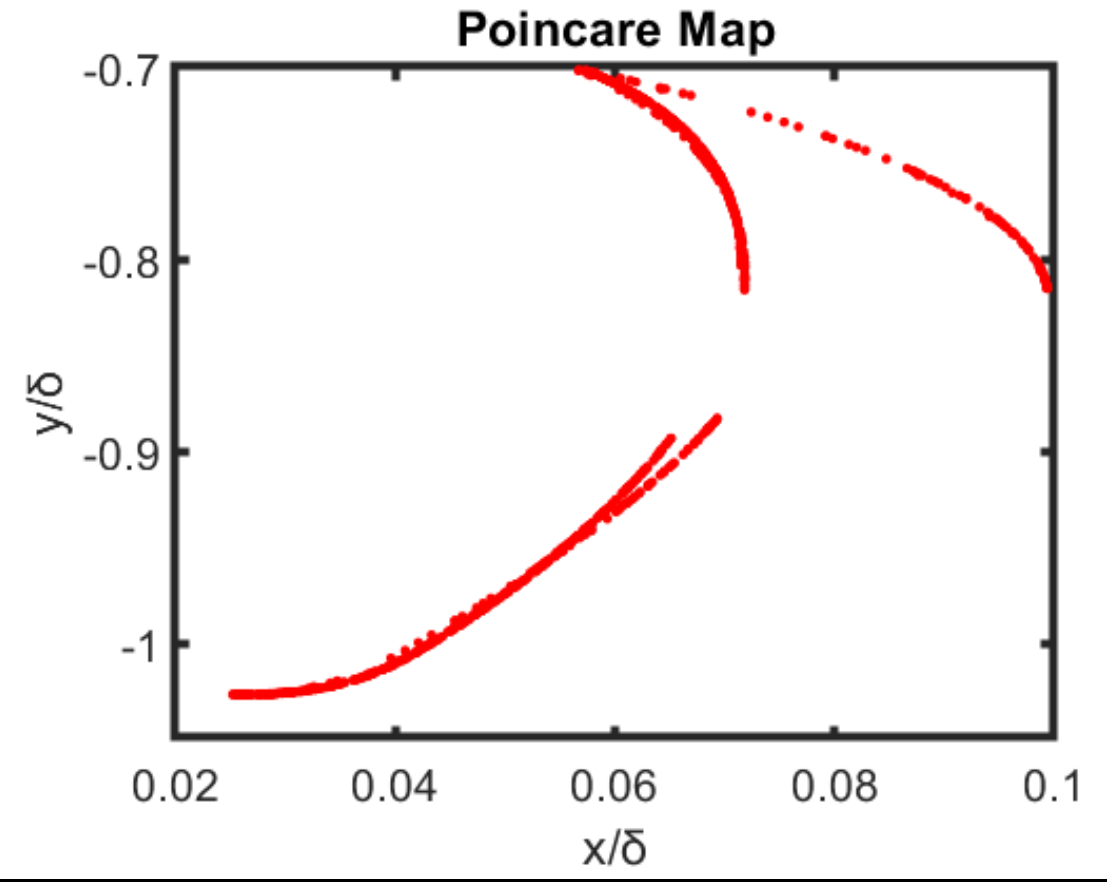

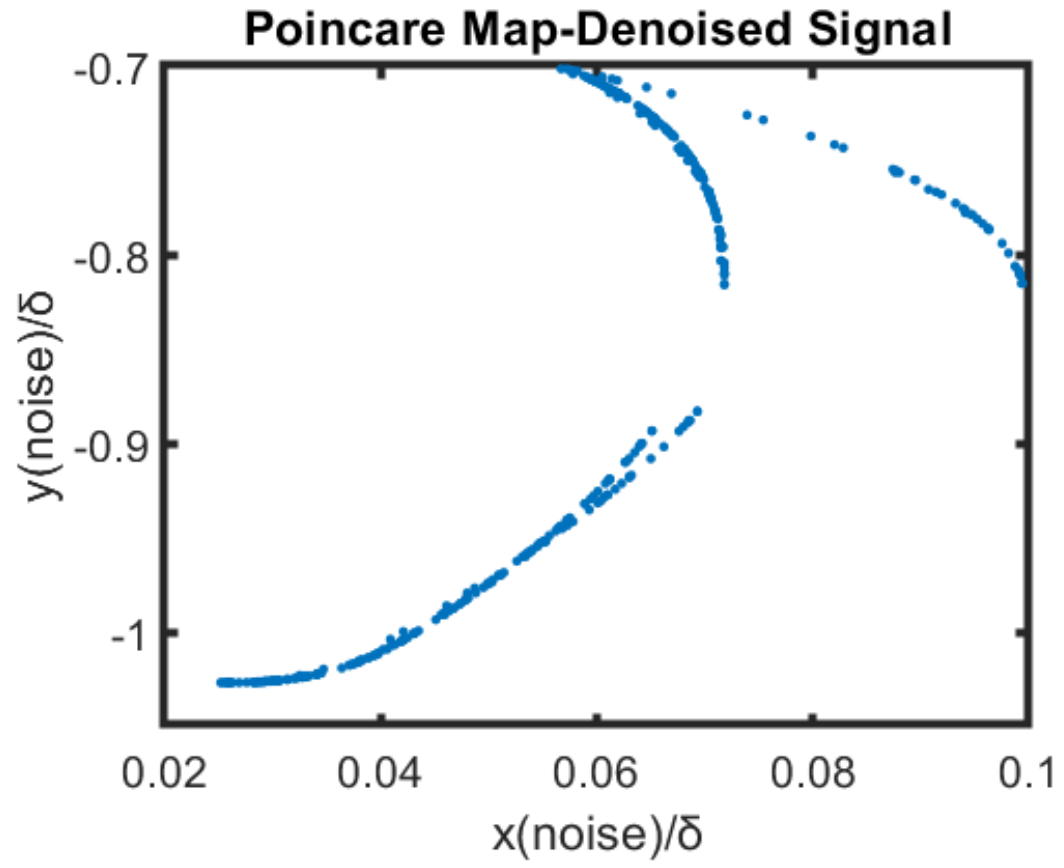

| (a) | (b) |
|---|---|

Fig 18. Poincare Map of chaotic system (case-7) (a) before applying noise reduction (b) after applying noise reduction

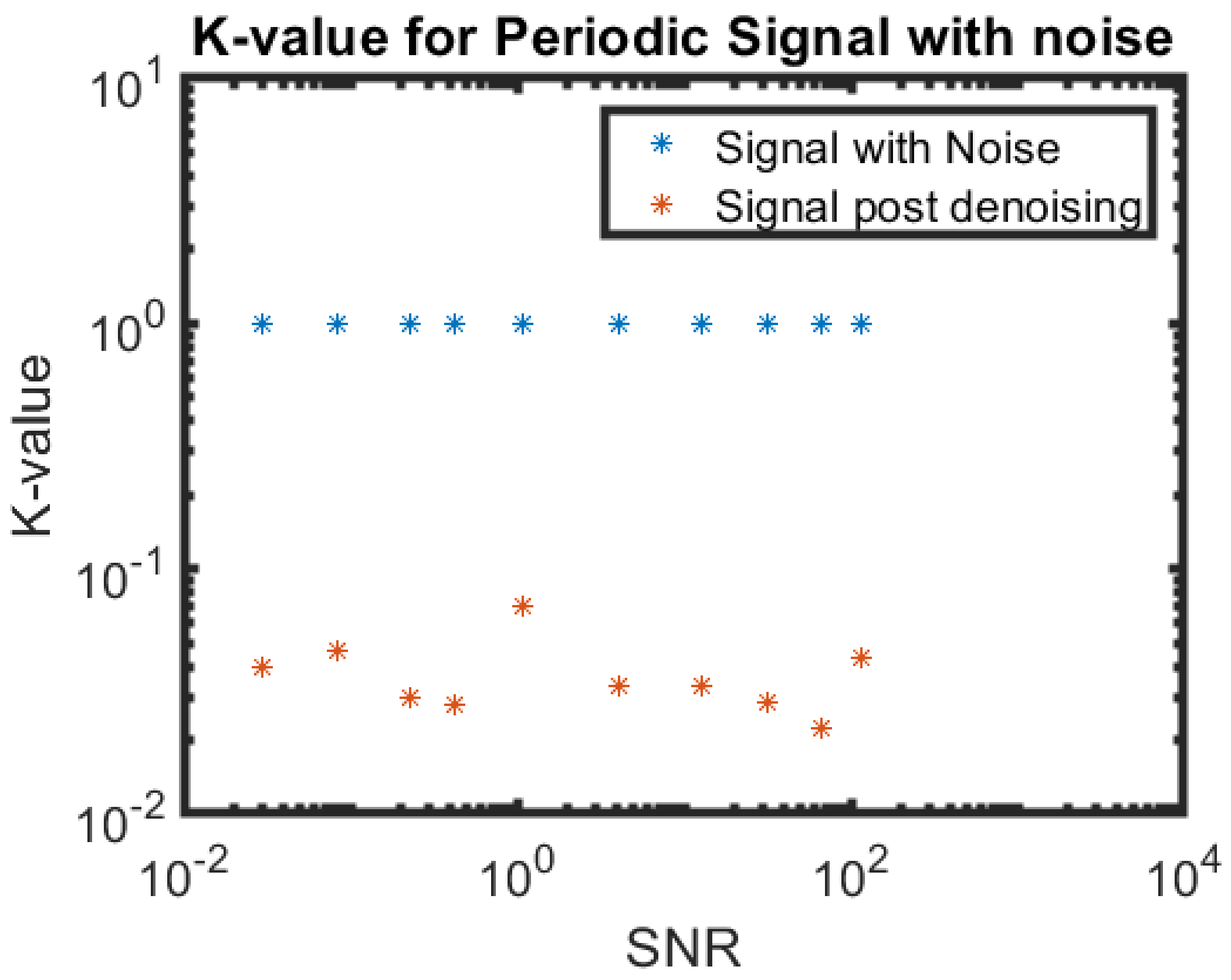

Figure 19. comparison of K-value from 0-1 test for periodic signal (case 1) with noise and signal post noise reduction

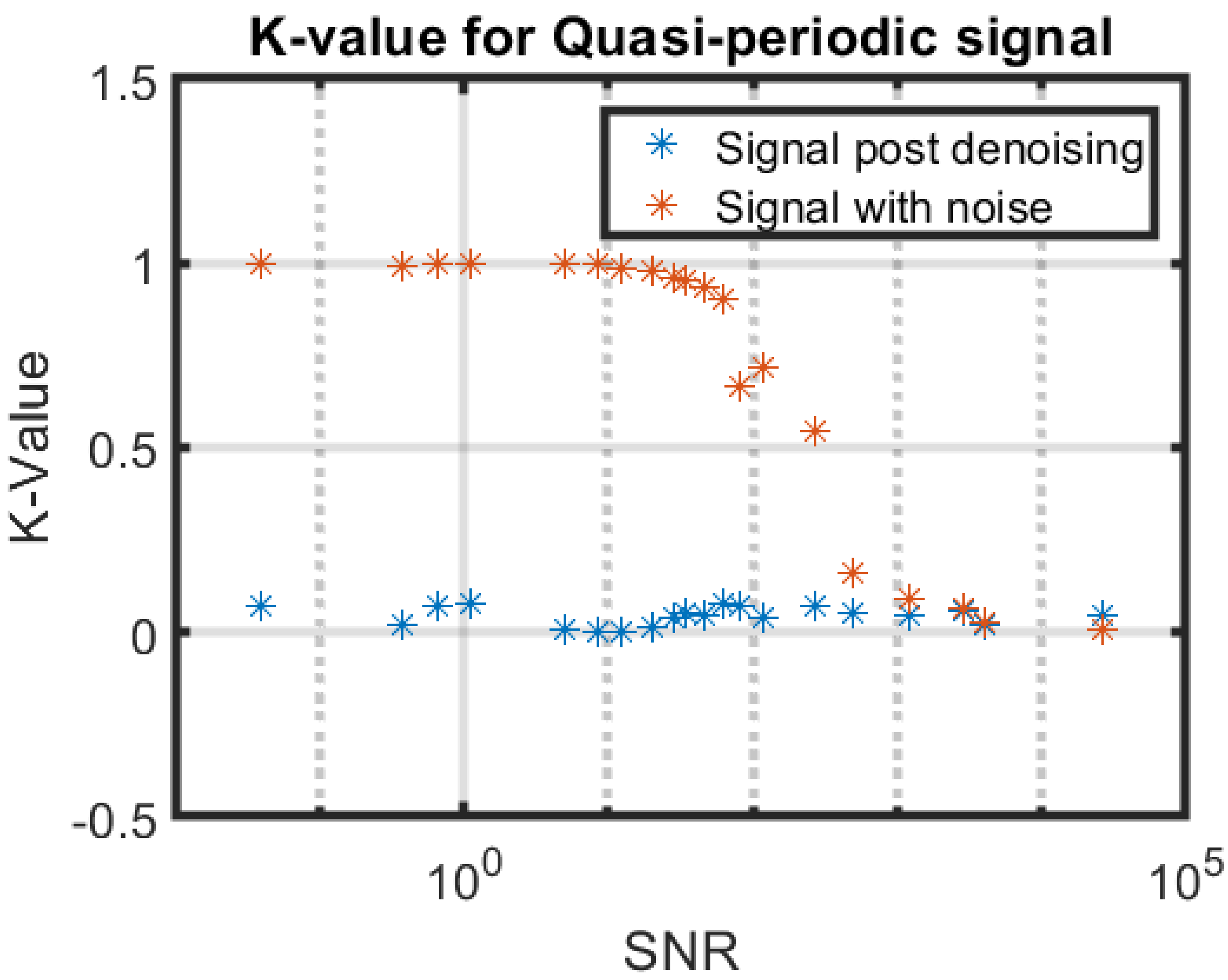

Figure 20. comparison of K-value from 0-1 test for quasi-periodic signal (case 8) with noise and signal post noise reduction

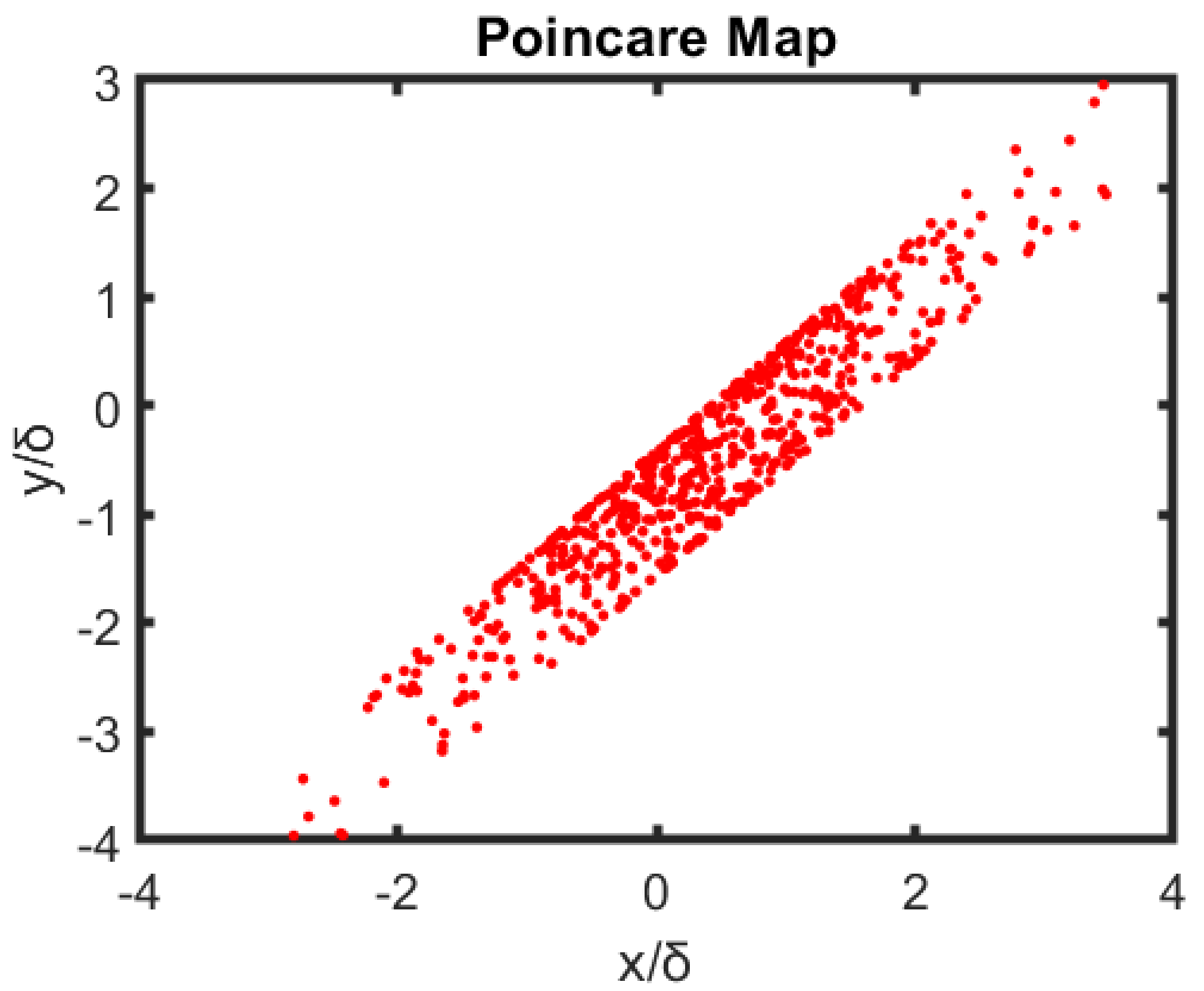

Figure 21 (a). Poincare Map of Quasi-periodic Vibration (case 8) with SNR 0.38

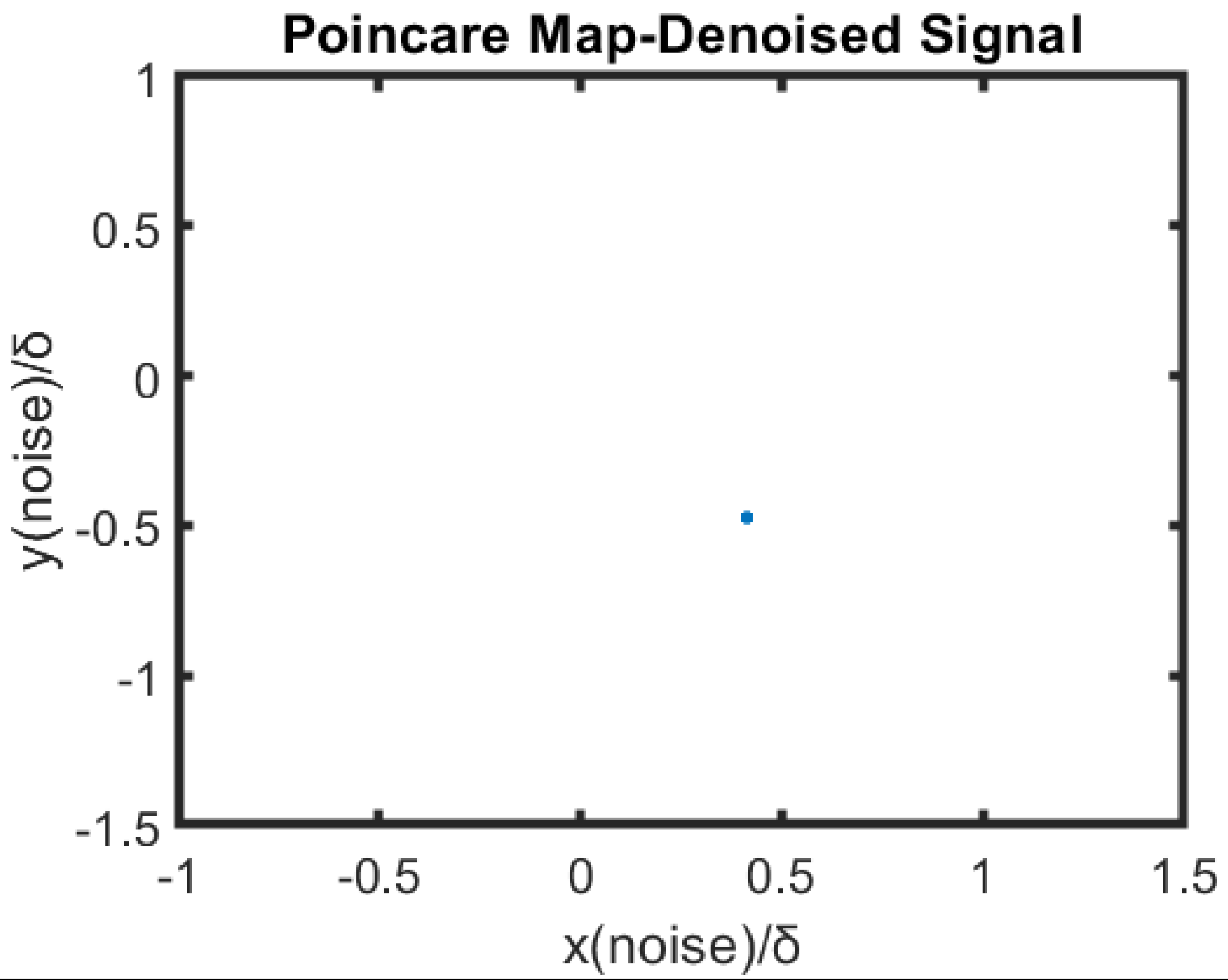

Figure 21 (b). Poincare Map of Quasi-periodic Vibration (case 8) with SNR 0.38 after noise reduction

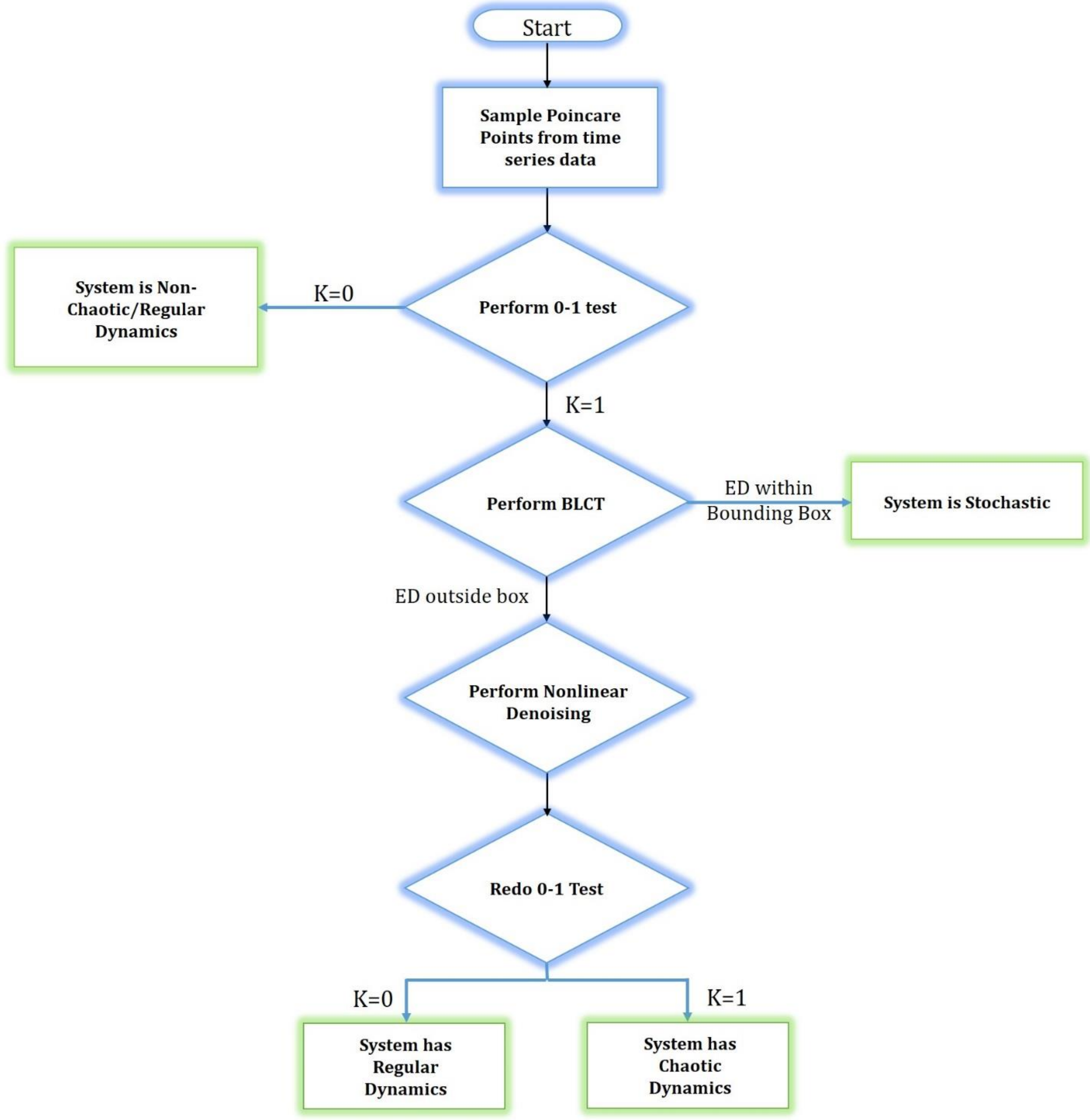

Fig 22. Decision Tree for Proposed Test for Chaos identification in Rotor-Stator Rub Model

Table 4. Proposed test on Signal mixed with Noise

| Primary Signal (From Table 1) | Noise Type-WGN Noise Level(SNR) | K Value | 0-1 Test Result(Signal Type) | K-value (Proposed Test) | Proposed test Result |
|---|---|---|---|---|---|
| Case 1 | 30 | 0.9990 | Chaotic | 0.042 | Regular Dynamics |
| Case 2 | 30 | 0.9976 | Chaotic | 0.05 | Regular Dynamics |
| Case 3 | 30 | 0.9969 | Chaotic | 0.02 | Regular Dynamics |
| Case 4 | 30 | 0.9987 | Chaotic | 0.033 | Regular Dynamics |
| Case 5 | 30 | 0.9980 | Chaotic | 0.031 | Regular Dynamics |
| Case 6 | 30 | 0.9985 | Chaotic | 0.04 | Regular Dynamics |
| Case 1 | 0.1 | 0.9981 | Chaotic | NA | Stochastic |
| Case 1 | 10 | 0.9995 | Chaotic | 0.022 | Regular Dynamics |
| Case 8 | NA | 0.9984 | Chaotic | 0.9988 | Chaotic |
| Case6 | 1 | 0.9988 | Chaotic | NA | Stochastic |